\documentclass[11pt]{article}
\usepackage{cite}
\usepackage{enumerate}
\usepackage[scale=0.8]{geometry}
\usepackage[utf8]{inputenc}

\usepackage{xcolor}
\usepackage{graphicx, graphbox}
\usepackage[subrefformat=parens,labelformat=parens]{subfig}
\usepackage{amsmath, amsfonts,amssymb,mathtools,nicefrac,nccmath,cases}
\usepackage{microtype,booktabs,multirow}
\usepackage{tikz}

\usepackage{amsfonts,amsmath}
\usepackage{bbm}
\usepackage{xspace}

\newcommand{\cP}{\mathcal{P}}
\newcommand{\cC}{\mathcal{C}}

\newcommand{\cK}{\mathcal{K}}
\newcommand{\cN}{\mathcal{N}}

\newcommand{\cH}{\mathcal{H}}

\newcommand{\cI}{\mathcal{I}}
\newcommand{\cJ}{\mathcal{J}}
\newcommand{\cR}{\mathcal{R}}

\newcommand{\cM}{\mathcal{M}}

\newcommand{\I}{\text{Im}}

\newcommand{\beq}{\begin{equation}}  \newcommand{\eeq}{\end{equation}}
\newcommand{\bal}{\begin{aligned}}   \newcommand{\eal}{\end{aligned}}
\newcommand{\bea}{\begin{eqnarray}}  \newcommand{\eea}{\end{eqnarray}}
\def\beqa{\begin{eqnarray}}
\def\eeqa{\end{eqnarray}}

\newcommand{\bmat}{\left(\begin{array}}
	\newcommand{\emat}{\end{array}\right)}

\DeclareMathOperator{\rk}{rk}

\usepackage{hyperref}

\begin{document}

\baselineskip=14pt
\parskip 5pt plus 1pt 

\vspace*{2cm}
\begin{center}
{\LARGE\bfseries Classifying Calabi-Yau threefolds \\[.3cm]
using infinite distance limits}\\[5mm]

\vspace{1cm}

{\bf Thomas W.~Grimm}${}^{\,a,}$\footnote{t.w.grimm@uu.nl},
{\bf Fabian Ruehle}${}^{\,b,c,}$\footnote{fabian.ruehle@cern.ch}
{\bf Damian van de Heisteeg}${}^{\,a,}$\footnote{d.t.e.vandeheisteeg@uu.nl},

{\small
\vspace*{.5cm}
${}^{a\;}$Institute for Theoretical Physics, Utrecht University\\ Princetonplein 5, 3584 CE Utrecht, The Netherlands\\[3mm]
${}^{b\;}$CERN Theory Department, 1 Esplanade des Particules, CH-1211 Geneva, Switzerland\\[3mm]
${}^{c\;}$Rudolf Peierls Centre for Theoretical Physics, University of Oxford\\ Parks Road, Oxford OX1 3PU, UK
}
\end{center}
\vspace{1cm}
\begin{abstract}\noindent
We present a novel way to classify Calabi-Yau threefolds by systematically studying their infinite volume limits. Each such 
limit is at infinite distance in K\"ahler moduli space and can be classified by an associated limiting mixed Hodge structure. 
We then argue that the such structures are labeled by a finite number of degeneration types that combine into a characteristic 
degeneration pattern associated to the underlying Calabi-Yau threefold. These patterns provide a new invariant way to present crucial 
information encoded in the intersection numbers of Calabi-Yau threefolds. For each pattern, we also introduce a Hasse diagram with vertices 
representing each, possibly multi-parameter, decompactification limit and explain how to read off properties of the Calabi-Yau manifold from 
this graphical representation.  In particular, we show how it can be used to count elliptic, K3, and nested fibrations and determine relations of elliptic fibrations under birational equivalence. We exemplify this for hypersurfaces in toric ambient spaces as well as for complete intersections in products of projective spaces. 
\end{abstract}

\newpage

\tableofcontents
\setcounter{footnote}{0}

%%%%%%%%%%%%%%%%%%%%%%%%%%%%%%%%%%%%%%%%%%%%%%%%%%%%%%%%%%%%%%%%%%%%%%%%%%%%
\newpage
\section{Introduction}
\label{sec:Introduction}
In the study of effective actions arising from string theory compactifications,  Calabi-Yau threefold backgrounds have been of 
interest for decades \cite{Candelas:1985en,Huebsch,Greene_StringonCY}. Compactifying 
the Type II or heterotic string theories on such Calabi-Yau manifolds leads to four-dimensional supergravity theories with $\cN=2$ or $\cN=1$ supersymmetry, respectively, while using them to compactify M-theory or F-theory yields five- or six-dimensional supergravity theories with minimal supersymmetry. It is a long-standing open problem to systematically classify the possible supergravity theories arising in these compactifications. In the spirit of the swampland program, one can 
then ask whether one can identify conditions on supergravity theories that need to be satisfied in order that they can be consistently coupled to a UV complete quantum gravity. In this work, we will make progress on this question by suggesting a new systematic way to classify Calabi-Yau manifolds, which implies a classification of the associated supergravity theories. Our approach rests on powerful mathematical results obtained 
in asymptotic Hodge theory. It continues recent efforts \cite{Grimm:2018ohb,Blumenhagen:2018nts,Lee:2018urn,Lee:2018spm,Grimm:2018cpv,Gonzalo:2018guu,Corvilain:2018lgw,Lee:2019tst,Font:2019cxq,Marchesano:2019ifh,Lee:2019xtm,Grimm:2019wtx,Erkinger:2019umg, Kim:2019vuc, Lee:2019skh,Demirtas:2019lfi, McNamara:2019rup,Kehagias:2019akr,Lee:2019wij,GLV} 
to use deep mathematical structures to test and extend the swampland conjectures about effective theories that are consistent with 
quantum gravity.

While many examples of Calabi-Yau threefolds are known, it is extremely hard to group them into equivalence classes that share common features. One way to approach this problem is to use Wall's theorem \cite{Wall:1966aaa}, which states that the homotopy types of Calabi-Yau threefolds are classified by the numerical data given by the Hodge numbers, the triple intersection numbers, and the divisor integrals of the second Chern class. However, this data is not easy to handle in practice. In fact, even checking whether or not two manifolds are homotopically equivalent can be a difficult task, since one needs to compare the triple intersections up to basis transformations. While certain basis independent invariants were identified in \cite{Huebsch}, these quantities capture only very limited information about the geometry and become increasingly weak for larger Hodge numbers. 

In this paper, we introduce a new classification using so-called limiting mixed Hodge structures~\cite{Schmid,CKS}. These structures arise in all limits at the boundaries of the complex structure and K\"ahler moduli space. Focusing on the complex structure moduli space of Calabi-Yau threefolds, they encode, roughly speaking, how the Hodge decomposition of the third  cohomology behaves at the boundaries of the moduli space. The construction proceeds by first associating to each boundary a nilpotent orbit \cite{Schmid}, 
which depends on the monodromy transformations and holomorphic data associated to the boundary component under consideration. Given a nilpotent orbit, 
one can then construct the associated limiting mixed Hodge structure. Crucially, such  mixed Hodge structures can be classified and hence used in 
a classification of possible degeneration limits \cite{Kerr2017} (see also \cite{Grimm:2018cpv}). We will discuss this classification in detail in Section \ref{LMH_sing}, where we 
also recall general rules for intersections of boundary components at which the degeneration of a Calabi-Yau threefold worsens. 

Using mirror symmetry, the classification of degeneration limits is also readily applied to the K\"ahler moduli space~\cite{Grimm:2018cpv,Corvilain:2018lgw}. 
In this case, it corresponds to a classification of all decompactification limits. The monodromy transformations are given in terms of the intersection numbers, while the additional holomorphic data on the boundary is fixed by specifying the integrated second Chern classes. 
The limiting mixed Hodge structure associated to a decompactification limit can be classified into $3h^{1,1}-1$ degeneration types denoted by II$_b$, III$_c$, and IV$_d$. We propose in Section~\ref{sec:EnhancementDiagrams} that any Calabi-Yau threefold can be associated with a corresponding enhancement pattern that can be determined by 
successively performing all possible decompactification limits. Note that the so-derived patterns naturally represent a partially ordered set, since 
they describe how the Hodge structure of the smooth threefold splits into finer and finer limiting mixed Hodge structure when sending more 
K\"ahler volumes towards the decompactification limit. It is therefore natural to associate with each Calabi-Yau threefold a Hasse diagram summarizing 
all its large volume degenerations. We will call these graphs in the following \textit{large volume enhancement diagrams} or, with the understanding that we are only at the large volume point in this work, simply \textit{enhancement diagrams}. This leads to a systematic grouping of all Calabi-Yau threefolds into equivalence classes and constitutes a novel classification. Recently, graphs~\cite{Apruzzi:2019vpe, Apruzzi:2019opn,Apruzzi:2019enx} and Hasse diagrams~\cite{Hassler:2019eso, Bourget:2019aer, Bourget:2019rtl,Hanany:2019tji} have been important tools in classifying five- and six-dimensional SCFTs.

The proposed classification possesses several interesting features, which we will only partly explore in this work. The enhancement graph can be used to determine whether two Calabi-Yau threefolds are homotopically inequivalent. In contrast to the invariants proposed in \cite{Huebsch}, this way of classifying threefolds becomes richer with increasing $h^{1,1}$. It is, however, important to stress that 
this classification is not fine enough to distinguish all Calabi-Yau threefolds, e.g.~certain rescalings of the intersection numbers will often not change the enhancement diagram. Nevertheless, we are able to demonstrate that the diagrams capture key features of the manifold, such as the presence of elliptic, K3, and nested fibrations and the relation of elliptic fibrations under birational equivalence.

We illustrate our finding using Calabi-Yau threefolds that are constructed as the anti-canonical hypersurface in a toric variety given by a reflexive polytope (as classified by Kreuzer and Skarke (KS)~\cite{Kreuzer:2000xy}), and by complete intersection Calabi-Yau manifolds (CICYs) in an ambient space that is given by a product of projective spaces~\cite{Candelas:1987kf,Anderson:2017aux}.

This paper is organized as follows. In Section \ref{LMH_sing}, we discuss how limiting mixed Hodge structures can be used to classify  degenerations of Calabi-Yau threefolds. We introduce the degeneration types that arise at large volume limits. In Section \ref{sec:EnhancementDiagrams} we then explain how taking successive limits leads to a characteristic enhancement pattern that can be captured by enhancement (or Hasse) diagrams. We explain how we treat Calabi-Yau threefolds with simplical as well as non-simplicical K\"ahler cones, give an example how geometric transitions via toric blowups lead to transitions among diagrams, and discuss symmetries and constraints of enhancement diagrams. In Section~\ref{sec:DiagramsFibrations}, we study Calabi-Yau manifolds from the KS and CICY list (up to $h^{1,1}=5$ and $h^{1,1}=10$, respectively), and illustrate how geometric properties are encoded in the enhancement diagrams. We illustrate how enhancement diagrams can be used to distinguish inequivalent Calabi-Yau threefolds in Section~\ref{sec:Classification}. In Section~\ref{sec:Conclusion}, we present our conclusions and give an outlook on future research directions involving these techniques. In the appendices, we give the necessary background on limiting mixed Hodge structure in Appendix~\ref{limitingMHS} and collect the enhancement diagrams from the KS and CICY scans for $h^{1,1}\leq3$ in Appendix~\ref{app:diagrams}.

%%%%%%%%%%%%%%%%%%%%%%%%%%%%%%%%%%%%%%%%%%%%%%%%%%%%%%%%%%%%%%%%%%%%%%%%%%%%
\section{Classifying degenerations using limiting mixed Hodge structures} \label{LMH_sing}

In this section we briefly summarize the mathematical results that are used to classify the limits in the K\"ahler moduli 
space of a Calabi-Yau threefold $Y_3$ largely following \cite{Grimm:2018cpv,Corvilain:2018lgw}. Note that these tools are more directly applicable for the variation 
of the Hodge $(p,q)$-decomposition of $H^3(Y_3,\mathbb{C})$ over the complex structure moduli space and are translated into the K\"ahler sector using mirror symmetry. The original results on the variation of Hodge structures \cite{Schmid,CKS} are abstract and more generally applicable.

\subsection{On the K\"ahler moduli space and decompactification limits}\label{class_infinite_distance_limits}

To begin with, we recall some basic facts about the K\"ahler moduli space
of a Calabi-Yau threefold $Y_3$. The K\"ahler structure is parametrized 
by the K\"ahler form $J$ on $Y_3$. The admissible K\"ahler forms are those that 
guarantee that the complex submanifolds of $Y_3$ have positive volume. Concretely 
this is ensured by the conditions
\beq \label{Kcone}
  \int_{\mathcal{C}} J > 0\; , \qquad \int_{D} J \wedge J > 0 \; , \qquad \int_{Y_3} J \wedge J \wedge J > 0 \; ,
\eeq
where $\mathcal{C}$, $D$ are holomorphic curves and divisors, respectively.\footnote{The closure of the K\"ahler
cone is the cone $\overline{\textit{NE}}^1(Y_3)$ of nef classes, which is dual to the closure of the Mori cone $\overline{\textit{NE}}_1(Y_3)$ of effective two-cycles.} 
Note that the conditions \eqref{Kcone} define the cone of admissible K\"ahler forms, which is 
known as the K\"ahler cone. This cone can be simplicial or non-simplicial depending on the considered $Y_3$. In the former case, it is generated by exactly $h^{1,1}(Y_3)$ linearly independent 
forms $\omega_I$, while in the latter case one is required to specify also linearly dependent forms 
to describe its edges. In this work, we will study both CYs with simplicial and non-simplicial K\"ahler cones. 

In order to proceed, we next expanded the K\"ahler form 
$J= v^I \omega_I$ in an integral basis of two-forms $\omega_I \in H^2(Y_3,\mathbb{Z})$ with real coefficients $v^I$. 
The basis $\omega_I$ can be chosen such that the $v^I$ span a simplicial subcone of the full K\"ahler 
cone. We will pick such a basis in the following and then consider all possible simplicial subcones whose union leads 
to the full K\"ahler cone. Note that this implies that when taking a limit in K\"ahler moduli space sending 
one or more $v^I \rightarrow \infty$, we have to specify in which simplicial subcone this limit is taken. 
In this work, we exploit the fact that all such limits can be classified using asymptotic Hodge theory. We 
will briefly introduce the necessary mathematical machinery next. A more complete introduction to the mathematics 
can be found in the reviews \cite{CattaniKaplanAster}. Application to Calabi-Yau manifolds, including a short introduction 
of the mathematics, can also be found in \cite{Grimm:2018cpv}.

Taking any limit $v^I \rightarrow \infty$ leads to a decompactification of the  Calabi-Yau manifold $Y_3$, which implies that we are approaching the boundaries of the moduli space and our standard tools of working with 
forms and integrals become obsolete. 
 This has a clear interpretation if we map limits in the K\"ahler moduli space to limits in the complex structure 
 moduli space by using mirror symmetry. In fact, one finds that in such limit the 
 Hodge structure of $H^{3}(Y_3,\mathbb{C})$, i.e.~the decomposition into $(p,q)$-forms, degenerates.
In order to apply the mirror map, one first has to complexify the K\"ahler volumes $v^I$ into complex 
coordinates $t^I = b^I + i v^I$. For string theory on $Y_3$, the scalars $b^I$ have the interpretation as 
modes of the $B_2$-field under which the string is charged. 
Mirror symmetry then exchanges the large volume regime $\I\, t^I \gg 1$, 
with the large complex structure regime by identifying $t^I$ with the complex structure deformations~$z^I$ of a mirror 
Calabi-Yau threefold $\tilde Y_3$. In the following, we will simultaneously talk about the large volume regime for $Y_3$ and the 
large complex structure regime for $\tilde Y_3$, keeping in mind that these two regimes are interchanged by mirror symmetry. 

The metric on the complex structure moduli space of the mirror Calabi-Yau threefold~$\tilde Y_3$ is the 
famous Weil-Petersson metric and can be obtained from the K\"ahler potential 
\begin{equation}
\label{Kcs}
K(z,\bar z)=-\log(i\bar \Pi^{\hat I} \eta_{\hat I \hat J}\Pi^{\hat J})\; .
\end{equation}
Here we have introduced the pairing $\eta_{\hat I \hat J}$ and the periods $\Pi^{\hat I}$, arising by expanding the holomorphic (3,0)-form~$\Omega$ in a real integral basis
$\gamma_{\hat I}$, $\hat I=1,\dots,2h^{2,1}(\tilde Y_3)+2$, via 
\begin{equation}
\label{Pics}
\Omega=\Pi^{\hat I} \gamma_{\hat I}\; , \quad \eta_{{\hat I} \hat J}=-\int_{\tilde Y_3}\gamma_{\hat I}\wedge \gamma_{\hat J}\; .
\end{equation}
Note that these expressions are evaluated in a certain basis $\gamma_{\hat I}$ of three-forms, and it will be crucial to pick an 
appropriate one to find a simple match with the K\"ahler moduli. In fact, one can show that there is 
a choice of basis such that the $(2h^{1,1}(Y_3)+2)$-dimensional period vector~$\mathbf{\Pi}$
takes the following form in the large complex structure regime $\I\, t^I \gg 1$: 
\begin{equation}\label{Kahler-p}
\mathbf \Pi \, (t^I) =
\begin{pmatrix}
1\\
t^I\\
\frac{1}{2}\cK_{IJK} t^J t^K + \frac{1}{2}\cK_{IJJ} t^J - c_I  \\
\frac{1}{6}\cK_{IJK} t^I t^J t^K - (\frac{1}{6} \cK_{III} + c_I) t^I + \frac{i \zeta(3) \chi}{8\pi^3}  
\end{pmatrix} \; .
\end{equation}
Here we have introduced the topological quantities 
\beq
   \cK_{IJK} = \int_{Y_3} \omega_I \wedge \omega_J \wedge \omega_K\; , \qquad  c_I = \frac{1}{24}\int_{Y_3} \omega_I \wedge c_2(Y_3) \; , \qquad \chi  = \int_{ Y_3} c_3( Y_3)\; ,
\eeq
where $\cK_{IJK}$ are the triple intersection numbers, and $c_2(Y_3)$, $c_3(Y_3)$ are the Chern classes of the tangent bundle of $Y_3$. 
Note that the expression \eqref{Kahler-p} can be derived by evaluating the so-called $\Gamma$-class 
in a certain K-theory basis for D6-, D4-, D2- and D0-branes wrapping the whole threefold, divisors $D_I$, curves $\cC^I$ and points in $Y_3$. 
The definition of the curves $\cC^I$ is non-trivial and is discussed for example in \cite{Gerhardus:2016iot}. The reader will also find a related 
discussion of \eqref{Kahler-p} in \cite{Bloch:2016izu}.

Note that the period vector \eqref{Kahler-p} transforms non-trivially  under the shift 
$b^A \rightarrow b^A + 1$ of any of the $h^{1,1}(Y_3)$ scalars $b^I$. 
This defines a monodromy transformation $T_A$ via  
\beq \label{monodromy-def}
 \mathbf{\Pi} (t^1,...,t^A+1,... ) = T_A^{-1} \, \mathbf{\Pi}(t^1,...,t^{A},...)\; . 
\eeq 
It turns out that the $T_A$ derived for the large complex structure periods \eqref{Kahler-p} 
are all unipotent. For each $T_A$ we can then define 
the log-monodromy matrix $N_A$ by setting
\begin{equation}\label{N=logT}
N_A = \log (T_A)\; ,
\end{equation}
which yields a nilpotent matrix. We will see in the following the 
these are key in the classification of 
limits in the Calabi-Yau moduli space. Using \eqref{Kahler-p}, they are readily determined to be
\begin{equation}\label{lcsNa} 
N_A  = \left(
\begin{array}{cccc}
0                   & 0                  & 0            & 0\\
-\delta_{AI}        & 0                  & 0            & 0\\
-\frac{1}{2} \cK_{AAI} & -\cK_{AIJ}           & 0            & 0\\
\frac{1}{6}\cK_{AAA}  & \frac{1}{2}\cK_{AJJ} & -\delta_{AJ} & 0
\end{array}\right)\; . 
\end{equation}
A second crucial ingredient is the pairing~$\eta$ introduced in \eqref{Pics}. In the special basis introduced above it takes the form
\begin{equation} \label{eta-lcs}
\eta = \left(
\begin{array}{cccc}
0                         & -\frac{1}{6} \cK_{JJJ} - 2c_J     & 0                & -1\\
\frac{1}{6}\cK_{III} + 2c_I & \frac{1}{2}(\cK_{IIJ} - \cK_{IJJ}) & \delta_{IJ}      &  0\\
0                         & -\delta_{IJ}                   & 0                &  0\\
1                         & 0                              & 0                &  0
\end{array}\right)\; .
\end{equation}
Note that the expression \eqref{Kahler-p} can be written in a particularly simple 
form as
\beq \label{nil-gen}
   \mathbf{\Pi}  = \text{exp} \Big(-\sum_{I} t^I N_I \Big) \mathbf{a}_0\; , \qquad  \mathbf{a}_0  = \left(
1 , 0 , - c_{I} , \frac{i \zeta(3) \chi}{8 \pi^3}\right)^T\; . 
\eeq
As we will discuss in the following, this is not a special feature of the large complex structure 
periods \eqref{Kahler-p}, but rather a consequence of a powerful theorem of \cite{Schmid} 
that associates a so-called nilpotent orbit to each limit in complex structure and hence K\"ahler moduli space.
Using \eqref{nil-gen}, it is immediate that $ \mathbf{\Pi}$ transforms as in \eqref{monodromy-def} when using \eqref{N=logT}.

Let us now consider a limit in which $n$ of the $h^{1,1}(Y_3)$ coordinates $t^I$ are send to constant real parts plus 
$i \infty$. These limits lead to a decompactification of $Y_3$ and we will henceforth call them degeneration limits. 
Of course, there are many different ways a degeneration limit can be taken and it turns out that we 
have to restrict to subsectors in the K\"ahler cone to classify these limits. For example,  a limit 
falls into a certain subsector if we fix an ordering in which the $t^I$ are sent to the limit. To be more concrete, let us introduce an index set $\cI = (i_1,...,i_n)$, which labels 
the coordinates $t^{i_k}$ that are send to the limit. For a given limit we can then identify 
a \textit{growth sector}. These sectors are defined by 
\begin{equation}\label{eq:growthsector}
\cR_\cI \equiv \cR_{i_1\cdots i_n} = \left\{\ t^{i_k}=b^{i_k}+iv^{i_k}\ \bigg| \ \frac{v^{i_1}}{v^{i_{2}}}> \lambda \, , \ \ldots ,  \ \frac{v^{i_{n-1}}}{v^{i_n}}> \lambda \, , \ v^{i_n} > \lambda\, , \quad b^i < \delta\,\right\} \; ,
\end{equation}
with positive $\delta>0$ and $\lambda \gg 0$. Roughly speaking the growth sector defines which of the coordinates 
grows the fastest, which is the second fastest and so on. It therefore defines an ordering in the set of coordinates sent to the limit.\footnote{Note that this defines a partial ordering on the set of growth sectors.} 
One example of a limit lying in \eqref{eq:growthsector} is an ordered limit, where we first send $v^{i_1} \to  \infty$, then $v^{i_{2}}\to  \infty$, up to $v^{i_n} \to  \infty$.
The $\cR_{\cI}$ cut out a subregion of the K\"ahler cone. Of course, we can always reorder the coordinates $t^I$, such that the considered limit corresponds to sending the first $n$ coordinates to $i \infty$ as done in \cite{Grimm:2018cpv,Corvilain:2018lgw}. However, as mentioned above, in the non-simplicial case the $t^I$ correspond to a specific simplicial subcone. The full K\"ahler cone is then obtained 
by gluing these subcones together and ensuring consistency of the limits in the various $\cR_{\cI}$. In these cases, we need to use different index sets $\cI$. In order to keep the discussion general, we will use index sets throughout this work.

Having introduced the limits $t^{i_k} \rightarrow i \infty$, one can now 
use a powerful result of asymptotic Hodge theory known as the nilpotent orbit theorem \cite{Schmid}.
It asserts that one can associate to each limit in complex structure moduli space a so-called 
nilpotent orbit $\mathbf{\Pi}_{\rm nil}$. This orbit approximates the periods up to exponentially 
suppressed corrections of order $e^{i t^{i_k}}$ in any of the variable taken to the limit. Importantly, 
for the large complex structure periods, the nilpotent orbit associated to a limit is easily read off from \eqref{nil-gen}
and reads
\beq
     \label{nil-limit}
   \mathbf{\Pi}_{\rm nil}^{[\cI]}  = \text{exp} \Big(-\sum_{i \in \cI} t^i N_i \Big) \mathbf{a}^{[\cI]}_0\; , 
\eeq
where $\mathbf{a}^{[\cI]}_0$ contains the $t^I$, $N_I$ that are not taken to the limit,
\begin{align}
[\cI]=(1,\ldots,2 h^{1,1}+2)~\backslash~\cI\; .
\end{align}
A second important 
result of \cite{Schmid,CKS} is that to each nilpotent orbit \eqref{nil-limit} one can associate 
a \textit{limiting polarized mixed Hodge structure}. We will define such structures in appendix \ref{limitingMHS}. 
For us, it is crucial that such structures can be classified, as we discuss next.

\subsection{Classifying infinite distance limits in K\"ahler moduli space}

To introduce the classification \cite{Kerr2017,Grimm:2018cpv}, we first note that 
we can associate a certain log-monodromy matrix $N_{(\cI)}$ to each limit, defined by\footnote{Let us stress that  any positive linear combination of the $N_{i_k}$ would work equally well \cite{CattaniKaplan}.}
\begin{equation}\label{gen-limit}
t^\cI \equiv (t^{i_1},...,t^{i_n}) \rightarrow i \infty \quad \longrightarrow \quad N_{(\cI)} = N_{i_1} + ...+ N_{i_n}\; ,
\end{equation}
where $\cI = (i_1,...,i_n)$ is an ordered index set specifying the growth sector \eqref{eq:growthsector}.
Note that each $t^{i_k}$ can have a constant real part in the limit, which we set to zero in the following.\footnote{The boundary component 
approached in the limit \eqref{gen-limit} is of complex co-dimension $n$.}
The association of log-monodromy matrices to a limit can actually be done 
for any degeneration limit in complex structure moduli space. The large complex structure 
limits, which are mirror to degeneration limits in K\"ahler moduli space, are thus only specific examples. 
It is therefore no extra effort to introduce the general classification of log-monodromy matrices for Calabi-Yau threefolds before returning to the large complex structure/large volume 
setting. Let us consider an $m$-dimensional complex structure moduli space. 
We also abbreviate the log-monodromy matrix associated to the considered degeneration limit by~$N$, rather than $N_{(\cI)}$.
The pairing between two three-forms is denoted by $\eta$ as in \eqref{Pics}.  
The allowed pairs~$(N,\eta)$ can be classified into~$4 m$ degeneration types 
denoted by 
\bea \label{sing_types}
&\text{I}_a\ ,  &\quad a=0,...,m\; , \nonumber \\
& \text{II}_b\ ,&\quad b=0,...,m-1 \; ,\\  
&\text{III}_c\ ,&\quad c= 0,...,m-2 \; ,\nonumber \\ 
&\text{IV}_d\ , & \quad d = 1,...,m \; . \nonumber 
\eea
One can now show that these degeneration types classify the limiting mixed Hodge structures that can arise at any  
limit in complex structure moduli space reaching its  boundaries. 
The types are distinguished~\cite{Grimm:2018cpv} by the conditions listed in Table~\ref{table1}, where we stress that the categorization of cases I$_a$ and II$_b$ needs both~$\eta$ and~$N$, while cases III$_c$ and IV$_d$ only depend on $N$.
\begin{table}[!ht]
	\centering
	\begin{tabular}{@{}l@{\hspace{20pt}}lcc@{\hspace{20pt}}r@{}}\toprule
		\multirow{2}{*}{Type} \hspace*{.5cm} & \multicolumn{3}{c}{\hspace*{-2.5em} rank of} & \multirow{2}{*}{eigenvalues of $\eta N$}  \\[1pt] %\cmidrule{2-4}
		&$N$ &$N^2$&$N^3$ & \\ \midrule
		I$_a$ &$a$ & 0& 0 &$a$ negative\\
		II$_b$ &$2+b$ & 0& 0 & 2 positive, $b$ negative\\
		III$_c$ &$4+c$ & 2& 0 & not needed\\
		IV$_d$ &$2+d$ & 2& 1 & not needed\\
		\bottomrule
	\end{tabular}
	\caption{Classification of pairs $(N, \eta)$ allowed at limits of the complex structure 
		moduli space of Calabi-Yau threefolds.}
	\label{table1}
\end{table}

Having introduced a classification of the possible degeneration types of $(N_{(\cI)},\eta)$ occurring 
at any limit \eqref{gen-limit}, we can now use this to perform successive limits. More precisely, 
we can successively send the $t^{i_{k}} \rightarrow i \infty$ for $k=1,...,n$ and record the occurring 
degeneration type at each step. Let us denote the degeneration type~\eqref{sing_types} that occurs at the~$k^\text{th}$ step by ${\sf Type\ A}_{ (i_k)}$. 
We then find what we call an \textit{enhancement chain} of the form
\begin{equation}\label{t>inf_chain}
\text{I}_0\xrightarrow{\ t^{i_1} \rightarrow i\infty\ }\  {{\sf Type\ A}_{(i_1)}}\ \xrightarrow{\ t^{i_2} \rightarrow i\infty\ }\  {\sf Type\ A}_{(i_2)} \ \xrightarrow{\ t^{i_3} \rightarrow i\infty\ }\  
... \ \xrightarrow{\ t^{i_n} \rightarrow i\infty\ }\  {\sf Type\ A}_{(i_n)} \; ,
\end{equation}
where $I_0$ represents the non-degenerate geometry. 
Remarkably, one can now show that there are various constraints on allowed enhancement chains \cite{Kerr2017} (see \cite{Grimm:2018cpv}
for a discussion focused on the Calabi-Yau threefold case). 
For example, one can show that the degeneration type can only
increase or stay the same. Hence, a general 
 enhancement chain always takes the form 
\beq
\text{I}_{0}  \rightarrow ...  \rightarrow\, \text{I}_{a_k} \rightarrow \, \text{II}_{b_1}  \rightarrow ...  \rightarrow\, \text{II}_{b_l}   \rightarrow
 \, \text{III}_{c_1}  \rightarrow ...  \rightarrow\, \text{III}_{c_p}  \rightarrow \, \text{IV}_{d_1}  \rightarrow ...  \rightarrow\, \text{IV}_{d_q}\; . 
\eeq
The full list of allowed enhancements can be found in Table \ref{allowed_enhancements}. 
Note that these conditions arise non-trivially from the fact that a \textit{polarizable} mixed Hodge 
structure is associated  to each degeneration type. 

\begin{table}[t]
\centering
\begin{tabular}{cl}
\toprule
    \rule[-.16cm]{0cm}{0.6cm} \qquad starting  type \qquad\qquad& enhanced  type \qquad \qquad \\
   \midrule 
    \multirow{4}{*}{I$_a$} 
    & \rule[-.2cm]{0cm}{0.8cm} I$_{\hat a}$ for $a\leq \hat a$ \\
    &\rule[-.2cm]{0cm}{0.6cm} II$_{\hat b}$ for $a \leq \hat b $, $a < m$ \hspace*{.4cm}\\
    &\rule[-.2cm]{0cm}{0.6cm} III$_{\hat c}$ for $a \leq \hat c$, $a < m$\\
    &\rule[-.4cm]{0cm}{0.8cm} IV$_{\hat d}$ for $a < \hat d$, $a < m$ \\
     \multirow{3}{*}{II$_b$} 
    &\rule[-.2cm]{0cm}{0.8cm} II$_{\hat b}$ for $b\leq \hat b$\\
    &\rule[-.2cm]{0cm}{0.6cm} III$_{\hat c}$ for $2 \leq b \leq \hat c + 2$\\
    &\rule[-.4cm]{0cm}{0.8cm} IV$_{\hat d}$ for $1 \leq b \leq \hat d-1$ \\      
     \multirow{2}{*}{III$_c$} 
    &\rule[-.2cm]{0cm}{0.8cm} III$_{\hat c}$ for $c\leq \hat c$ \\
    &\rule[-.4cm]{0cm}{0.8cm} IV$_{\hat d}$ for $c + 2 \leq \hat d$ \\ 
       IV$_d$ & \rule[-.4cm]{0cm}{1cm} IV$_{\hat d}$ for $d\leq \hat d$  \\
    \bottomrule
\end{tabular}
\begin{picture}(0,0)
\put(-183,65){\begin{tikzpicture}
\draw [->] (0,0) -- (2,.7);
\end{tikzpicture}}
\put(-183,65){\begin{tikzpicture}
\draw [->] (0,0) -- (2,.1);
\end{tikzpicture}}
\put(-183,52){\begin{tikzpicture}
\draw [->] (0,-.2) -- (2,-.66);
\end{tikzpicture}}
\put(-183,38){\begin{tikzpicture}
\draw [->] (0,-.2) -- (2,-1.15);
\end{tikzpicture}}
%%%%%
\put(-183,-7){\begin{tikzpicture}
\draw [->] (0,0) -- (2,.5);
\end{tikzpicture}}
\put(-183,-10){\begin{tikzpicture}
\draw [->] (0,0) -- (2,-.1);
\end{tikzpicture}}
\put(-183,-26){\begin{tikzpicture}
\draw [->] (0,0) -- (2,-.66);
\end{tikzpicture}}
%%%%%%
\put(-183,-62){\begin{tikzpicture}
\draw [->] (0,0) -- (2,.25);
\end{tikzpicture}}
\put(-183,-71){\begin{tikzpicture}
\draw [->] (0,0) -- (2,-.32);
\end{tikzpicture}}
%%%%
\put(-181,-102){\begin{tikzpicture}
\draw [->] (0,0) -- (2,0);
\end{tikzpicture}}
\end{picture}
\caption{List of all allowed enhancements of degeneration types \cite{Kerr2017}, where $m$ is the dimension of the 
moduli space.}
\label{allowed_enhancements}
\end{table}

It is important to stress that the constraints of Table \ref{allowed_enhancements} only restrict the form of 
an enhancement chain, but do not yet cover all rules specifying which types are compatible when considering all possible 
limits. For example, let us denote by ${\sf Type\ A}_{a}$ and ${\sf Type\ A}_{b}$  the types occurring when sending 
$t^{a} \rightarrow i \infty$ and 
$t^{b} \rightarrow i \infty$, respectively. Clearly, one can also consider sending both $t^a,t^b \rightarrow i\infty$
yielding a type denoted by ${\sf Type\ A}_{a+b}$. The interesting question is then which combinations of 
${\sf Type\ A}_{a}$, ${\sf Type\ A}_{b}$, and ${\sf Type\ A}_{a+b}$ are allowed and ensure the existence 
of a polarized mixed Hodge structure. These rules are not yet known, but first results and 
a study of specific examples can be found in \cite{Kerr2017}.

Having discussed the general classification, let us now return to the large complex structure and large volume regime $\I\, t^I \gg 1$ 
and discuss the structures arising in the possible limits. 
Firstly, using~\eqref{lcsNa}, it is straightforward to 
determine $N_{(\cI)}$ in terms of the intersection numbers as
\begin{equation}\label{Nn}
N_{(\cI)}  = \left(
\begin{array}{cccc}
0               &             0             &         0          & 0 \\
-\sum_{i \in \cI} \delta_{iI}      &             0             &         0          & 0 \\
-\frac{1}{2} \sum_{i \in \cI}\cK_{iiI} &       -\sum_{i \in \cI}\cK_{iIJ}       &         0          & 0 \\
\frac{1}{6}\sum_{i \in \cI}\cK_{iii}  & \frac{1}{2} \, \sum_{i \in \cI}\cK_{iJJ} & -\sum_{i \in \cI}\delta_{iJ} & 0
\end{array}\right) \; .
\end{equation}
In this case it is not hard to show by using~\eqref{lcsNa},~\eqref{eta-lcs} together with the fact 
that~$\cK_{IJK}\geq 0$ for a simplicial subcone of the K\"ahler cone, 
that the case I$_a$ actually does not arise in this regime. In fact, one can show \cite{Corvilain:2018lgw} that 
all limits \eqref{gen-limit} in K\"ahler moduli space are of infinite distance in the metric derived from~\eqref{Kcs} using \eqref{Kahler-p}, \eqref{eta-lcs}. 
The degeneration types I$_a$ are at finite distance and arise, for example, at the conifold point in complex structure moduli space. 
For the remaining three cases, the degeneration type of the individual limits \eqref{gen-limit} 
is evaluated by determining the ranks of $(N_{(\cI)},N^2_{(\cI)},N^3_{(\cI)})$. 
We first define 
\begin{equation}\label{notation}
\cK_{IJ}^{(\cI)} \equiv \sum_{i \in \cI} \cK_{iIJ}\; , \qquad \cK_{I}^{(\cI)}  \equiv \sum_{i,j \in \cI} \cK_{ijI}  \qquad \text{and} \qquad \cK_{}^{(\cI)}  \equiv \sum_{i,j,k \in \cI} \cK_{ijk}\;.
\end{equation}
The powers of $N_{(\cI)}$ are then computed to be 
\begin{equation}
N_{(\cI)}^2 =\left( \begin{array}{cccc} 0     &     0     & 0 & 0 \\
0     &     0     & 0 & 0 \\
\cK_{I}^{(\cI)} &     0     & 0 & 0 \\
0     & \cK_{J}^{(\cI)} & 0 & 0
\end{array}\right)\; , \qquad 
N_{(\cI)}^3  = \left(
\begin{array}{cccc}
0      & 0 & 0 & 0 \\
0      & 0 & 0 & 0 \\
0      & 0 & 0 & 0 \\
- \cK{}^{\cI} & 0 & 0 & 0
\end{array}\right) \; .
\end{equation}
It is now straightforward to use Table~\ref{table1} and translate the rank conditions into conditions on the intersection numbers.
The result is presented in Table~\ref{Type_Table2}.
\begin{table}[!ht]
	\centering
	\begin{tabular}{@{}lccr@{}} \toprule
		Type &~$\rk \cK_{}^{(\cI)}$ &~$\rk \cK_{I}^{(\cI)}$ &~$\rk \cK_{IJ}^{(\cI)}$ \\ \midrule 
		II$_b$ & 0 & 0 &~$b$ \\  
		III$_c$ & 0 & 1 &~$c+2$ \\
		IV$_d$ & 1 & 1 &~$d$ \\ \bottomrule
	\end{tabular}
	\caption{List of types in the large volume regime in the limit $t^{\cI}=(t^{i_1},...,t^{i_n}) \rightarrow i \infty$.
	        For numbers and vectors, we define the ranks~$\text{rk}(K_{}^{(\cI)})$ and~$\text{rk}(\cK_{I}^{(\cI)})$ to be either~$0$ or~$1$, depending on
		whether $\text{rk}(K_{}^{(\cI)})=0$ and $\cK_{I}^{(\cI)}=0~\forall~I$.}
	\label{Type_Table2}
\end{table}

This concludes our introduction of the classification of limits in the $h^{1,1}(Y_3)$-dimensional K\"ahler moduli space. The aim of the next sections 
is to argue that, when collecting the information about all possible limits, we obtain a classification of Calabi-Yau threefolds that captures many core features of the geometry. In particular, we show that the fibration structure can be inferred from the enhancement pattern in Section~\ref{sec:DiagramsFibrations}. 

%%%%%%%%%%%%%%%%%%%%%%%%%%%%%%%%%%%%%%%%%%%%%%%%%%%%%%%%%%%%%%%%%%%%%%%%%%%%
\section{Enhancement diagrams to classify CYs}\label{sec:EnhancementDiagrams}
Having introduced the relevant mathematical background  to classify degeneration limits in the K\"ahler moduli space in Section \ref{LMH_sing}, we now use these techniques to classify Calabi-Yau threefolds themselves. We systematically consider different paths that can be taken for such limits in the K\"ahler cone, and then characterize each Calabi-Yau threefold $Y_3$ via the resulting pattern of degenerations. We call these patterns \textit{enhancement patterns}, since they encode all enhancement chains \eqref{t>inf_chain} that can occur around the large volume point for the Calabi-Yau manifold $Y_3$. Thereafter, we construct a Hasse diagram from this data, which provides a natural graphical representation of these enhancement patterns in terms of graphs, dubbed \textit{enhancement diagrams}. These graphs serve as an invariant based on the intersection numbers $\cK_{IJK}$ of the Calabi-Yau threefold, cf.~Section \ref{sec:Classification}, and can be used to read off various properties of Calabi-Yau threefold as discussed for fibrations in Section \ref{sec:DiagramsFibrations}.

In order to be able to classify a Calabi-Yau threefold based on degeneration limits, we need to consider all regions of its K\"ahler moduli space entering such limits. This  moduli space has the structure of a cone, spanned by generators $\omega_I \in H^2(Y_3,\mathbb{R})$, such that the K\"ahler form $J=v^I \omega_I$ is valued inside the cone $v^I \geq 0$. As discussed above, these cones can be \textit{simplicial} or \textit{non-simplicial}.

\subsection{Enhancement diagrams for simplicial K\"ahler cones}
\begin{figure}[t]
\centering
\includegraphics[width=4cm]{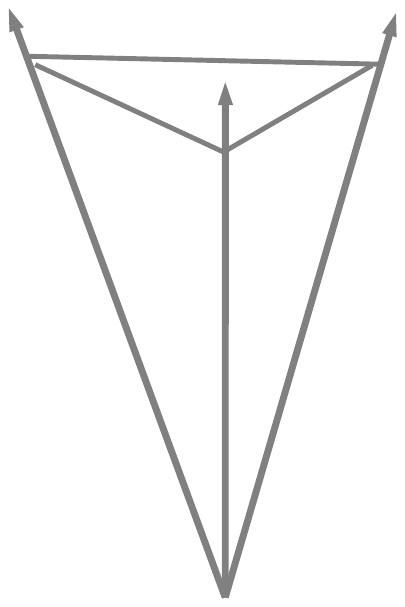} 
\begin{picture}(0,0)
\put(-69,128){$\omega_1$}
\put(-125,155){$\omega_2$}
\put(-8,155){$\omega_3$}
\end{picture}
\hspace{2.5cm}
\includegraphics[width=4cm]{simplicialcone.pdf} 
\begin{picture}(0,0)
%\put(245,170){$\omega_1$}
%\put(190,180){$\omega_2$}
%\put(290,180){$\omega_3$}
\put(-70,70){\tiny II$_1$}%1
\put(-95,70){\tiny III$_1$}%2
\put(-32,70){\tiny IV$_2$}%3
\put(-80,100){\tiny IV$_3$}%1,2
\put(-45,100){\tiny IV$_2$}%1,3
\put(-72,132){\tiny IV$_3$}%2,3
\put(-67,148){\tiny IV$_3$}%1,2,3
\end{picture}
\caption{Example for a simplicial K\"ahler cone with generators $\omega_i$, $i=1,2,3$. All faces of a simplicial K\"ahler cone are labeled by the degeneration types associated to the corresponding growth sectors, following enhancement pattern \eqref{eq:patternexample}.}\label{fig:simplicialexample}
\end{figure}

We begin our discussion with simplicial K\"ahler cones, and thereafter extend our discussion to non-simplicial cones in Section~\ref{sec:NonSimplicaialKCEnhancementDiagrams}. The decompactification limits correspond to limits $t^i \to i \infty$ for the K\"ahler moduli, where we will consider sending any (sub)set $\cI$ of these moduli to infinity. Following Table~\ref{Type_Table2}, one can label each of these limits based on the rank properties of the intersection numbers. More precisely, for a limit involving the moduli $t^i$, with $i \in \cI$, one computes the quantities $\rk \cK^{(\cI)}, \rk \cK^{(\cI)}_I,\rk \cK^{(\cI)}_{IJ}$, and deduces the corresponding degeneration type. When only a single modulus $t^i$ is send to $i \infty$, this means we compute $\rk \cK_{iii}, \rk \cK_{iiI}, \rk \cK_{iIJ}$, and associate the ray spanned by $(\omega_i)$ with the corresponding type of degeneration. Similarly, for two moduli $t^i,t^j$, we can label the face spanned by $(\omega_i,\omega_j)$ with the degeneration type. Continuing in this fashion, one can label all faces of the K\"ahler cone, as done for instance in Figure \ref{fig:simplicialexample}. 

Collecting all these sets of generators paired with degeneration types, we obtain the \textit{enhancement pattern} of the Calabi-Yau threefold at its large volume point. Namely, using this pattern one can deduce all possible manners in which the degeneration type can worsen by sending additional K\"ahler moduli to their limit, resulting in all \textit{enhancement chains} that occur around the large volume point. For instance, one could have
\begin{align}\label{eq:patternexample}
&\big((),\mathrm{I}_0 \big), \big((\omega_1) ,\mathrm{II}_1 \big),\big( (\omega_2) , \text{III}_1\big),\big( (\omega_3) , \text{IV}_2 \big),\nonumber\\
&\big( (\omega_1,\omega_2) , \text{IV}_3 \big)
, \big((\omega_1,\omega_3) , \text{IV}_2 \big), \big((\omega_2,\omega_3) , \text{IV}_3 \big), \big( (\omega_1,\omega_2,\omega_3) , \text{IV}_3 \big) \; .
\end{align}
And as an example, by first sending $t^1\to i\infty$, then $t^3 \to i \infty$, and lastly $t^2 \to i \infty$, we read off the enhancement chain
\begin{equation}\label{eq:chain}
\mathrm{I}_0 \xrightarrow{\ t^1 \rightarrow i\infty\ }\ \mathrm{II}_1 \xrightarrow{\ t^3 \rightarrow i\infty\ }\ \mathrm{IV}_2 \xrightarrow{\ t^2 \rightarrow i\infty\ }\ \mathrm{IV}_3\; ,
\end{equation}
from the enhancement pattern \eqref{eq:patternexample}, via the limits associated with $(\omega_1), (\omega_1,\omega_3)$ and $(\omega_1,\omega_2,\omega_3)$ respectively.

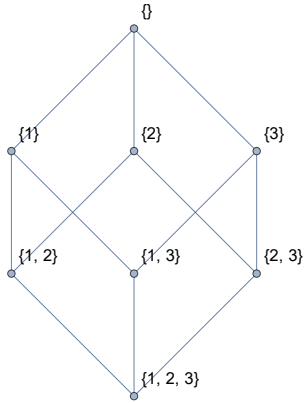
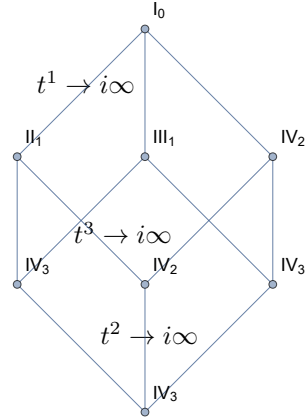
\begin{figure}[t]
\centering
\subfloat[Hasse diagram.]{\label{fig:hassediagram}\includegraphics[width=4cm]{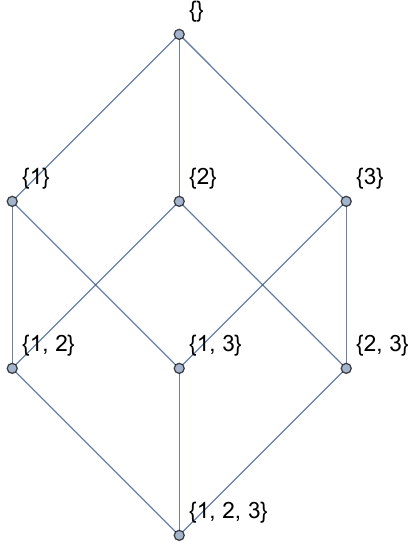}} 
\hspace{3cm}
\subfloat[Corresponding enhancement diagram.]{\label{fig:hassediagramexample}%
\hspace{10mm}%
\begin{picture}(0,0)%
\put(11,125){\small $t^1\to i\infty$}%
%\put(225,190){\tiny $t^2\to i\infty$}
%\put(260,175){\tiny $t^3\to i\infty$}
%
%\put(160,110){\tiny $t^2\to i\infty$}
\put(25,68){\small $t^3\to i\infty$}%
%
%\put(190,60){\tiny $t^3\to i\infty$}
\put(35,30){\small $t^2\to i\infty$}%
\end{picture}%
\includegraphics[width=4cm]{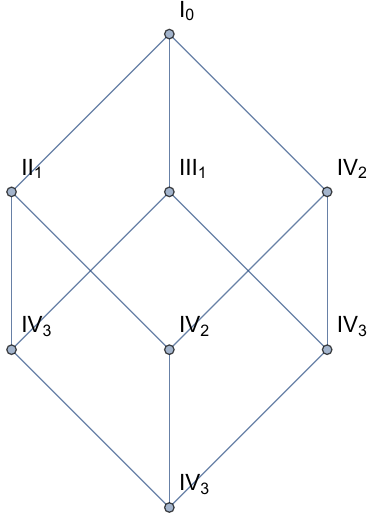}\hspace{10mm}
}
\caption{Example of an enhancement diagram of a CY with simplicial K\"ahler cone (polytope 230 in the Kreuzer-Skarke list with its unique fine regular star triangulation) and the corresponding enhancement diagram. The example is for the enhancement pattern of \eqref{eq:patternexample} with enhancement chain \eqref{eq:chain} explicitly indicated.}
\label{fig:exampleSimplicialKCDiagrams}
\end{figure}

To provide a natural way to present these enhancement patterns, we will make use of so-called \textit{Hasse diagrams}. Such a diagram simply displays a finite partially ordered set as a graph, via a drawing of its transitive reductions. Namely, one considers all elements in the set as vertices, and draws an edge between two elements $x,y$ if they satisfy $x<y$ and if there is no $z$ such that $x<z<y$. For our purposes, we will consider the power set of the set of generators of the K\"ahler cone as a finite partially ordered set, where inclusion serves to define a partial ordering of different elements of the power set, see for example in Figure~\ref{fig:hassediagram}.

Instead of labeling the vertices of the Hasse diagram with their associated generators, we label them with their degeneration type as obtained from the enhancement patterns, as done for instance in Figure~\ref{fig:hassediagramexample}. Then edges between vertices indicate one-step enhancements between limits, i.e.~sending an additional K\"ahler modulus to its limit. This results in a diagram consisting of $h^{1,1}+1$ rows, where we will count rows from 0 to $h^{1,1}$, such that the $n^\text{th}$ row corresponds precisely to a vertex for a set of $n$ generators. Note that this provides a convenient tool to read off all possible enhancement chains, since one can simply take all possible downward paths in the diagram. We will therefore call these diagrams \textit{enhancement diagrams}. Each enhancement chain always starts at I$_0$, where all K\"ahler parameters are at a generic point away from the boundary and the Calabi-Yau threefold has not yet degenerated. They also all end at the maximal degeneration type $\mathrm{IV}_{h^{1,1}}$ where all K\"ahler moduli are sent to their limit $t^1,\ldots,t^{h^{1,1}} \to i \infty$ \cite{Grimm:2018cpv,Corvilain:2018lgw}. 

The enhancement pattern can also be recovered straightforwardly from the enhancement diagram, up to a relabeling of the generators, as follows. The first row of the Hasse diagram contains all the limits associated with a single generator of the K\"ahler cone. Labeling each of the vertices in this row by a different generator $\omega_i$, we retrieve the limits of the enhancement pattern associated with a single generator. The sets of generators associated to lower vertices is determined by their connection to the vertices higher up in the enhancement chain. After including the non-degenerate phase $\mathrm{I}_0$, this reproduces the enhancement pattern. This strategy can also be applied to the enhancement diagrams of non-simplicial K\"ahler cones, as we discuss next.

\subsection{Example: Enhancement diagrams for geometric transitions of the quintic}
\label{sec:QuinticExample}
\begin{figure}[t]
\centering
\begin{minipage}[b]{0.15\textwidth}
\includegraphics[align=c,height=3.5cm]{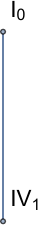}\vspace*{1.3cm}
\end{minipage}~\qquad
\begin{minipage}[b]{0.25\textwidth}
\centering
\includegraphics[align=c,height=5cm]{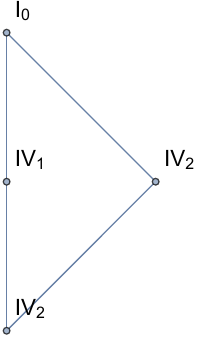}\vspace*{1cm}

\includegraphics[align=c,height=5cm]{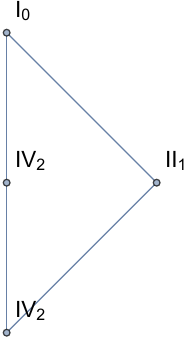}
\end{minipage}~\qquad
\begin{minipage}[b]{0.35\textwidth}
\centering
\includegraphics[align=c,height=6cm]{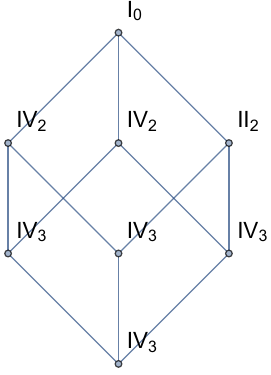}\vspace*{0.1cm}
\end{minipage}

\begin{picture}(0,0)
\put(-170,180){\rotatebox{20}{$\xrightarrow{\ \ \text{dP$_6$} \ \ }$}}
\put(-170,120){\rotatebox{-20}{$\xrightarrow{\ \text{conifold}\ }$}}

\put(-10,105){\rotatebox{20}{$\xrightarrow{\ \ \text{dP$_6$} \ \ }$}}
\put(-10,195){\rotatebox{-20}{$\xrightarrow{\  \text{conifold}\ }$}}
\end{picture}
\caption{Enhancement diagram of the quintic, a dP$_6$ transition for the quintic, a conifold transition for the quintic, and a combined dP$_6$ and conifold transition for the quintic.}
\label{fig:HasseDiagramsQuinticChain}
\end{figure}
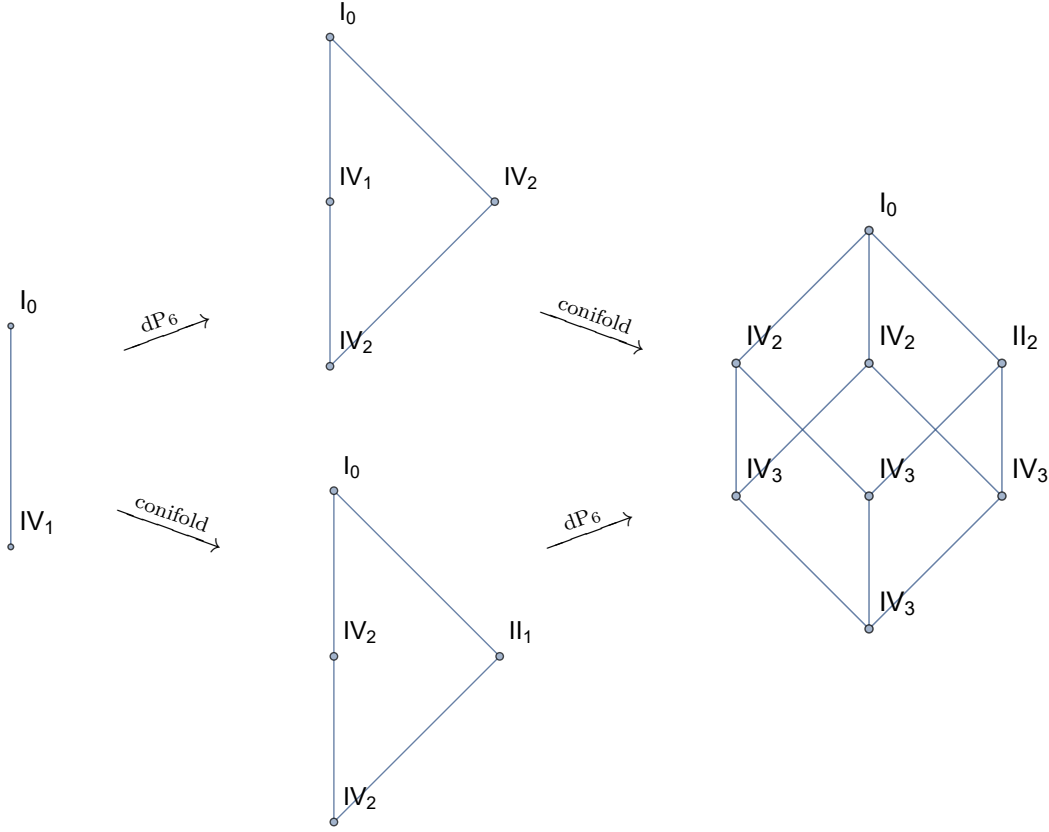
In order to illustrate how the enhancement diagrams change, we study two types of toric transitions starting from the quintic. The transitions for the diagrams are shown in Figure \ref{fig:HasseDiagramsQuinticChain}. The toric realization of these transitions have been described in~\cite{Grimm:2008ed}.

The quintic is given as the anticanonical hypersurface inside $\mathbbm{P}^4$, which can be described by a reflexive polytope with vertices 
\begin{align}
v_1=(1,1,1,1)\; ,~v_2=(-1,0,0,0)\; ,~v_3=(0,-1,0,0)\; ,~v_4=(0,0,-1,0)\; ,~v_5=(0,0,0,-1)\; .
\end{align}
Each of these vertices corresponds to a toric coordinate $x_i$, whose zero locus gives rise to a toric divisor $D_i=\{x_i=0\}$. They are all linearly equivalent and correspond to the hyperplane divisor of $\mathbbm{P}^4$. Since the quintic has $h^{1,1}=1$, it is favorable and a basis of $H^{1,1}(X)$ is given by pulling back the hyperplane class to $X$. The triangulation in the geometric phase is unique, and the Stanley-Reisner ideal is simply
\begin{align}
\text{SRI}=\langle x_1 x_2 x_3 x_4 x_5\rangle\; .
\end{align} 
The K\"ahler cone is simplicial and spanned by any of the (linearly equivalent) $D_i$ (or rather, the two-forms dual to the divisors), so that the K\"ahler form is given by
\begin{align}
\label{eq:KCQunitic}
J=t_1 J_1\; ,
\end{align}
with $J_1=D_1$. The infinite distance limit $t_1\to i\infty$ corresponds to a decompactification of the entire CY and is of type IV$_1$.

By adding a vertex
\begin{align}
\label{eq:KCQuniticdP6}
v_6=(0,0,0,1)
\end{align}
to the polyhedron, we restrict the dual polyhedron such that a dP$_6$ singularity occurs in the quintic hypersurface. The triangulation is still unique and the K\"ahler cone is still simplicial and can be written as
\begin{align}
J=t_1 J_1+t_2 J_2\;,
\end{align}
with $J_1=D_5$ and $J_2=D_5-D_6$. The limit $t_1\to i\infty$ still gives rise to a IV$_{1}$ degeneration, while $t_2\to i\infty$ as well as the combined limit $t_1,t_2\to i\infty$ give rise to a IV$_{2}$ degeneration.

If we add instead of $v_6$ a vertex
\begin{align}
v_6^\prime=(0,1,1,1)\; ,
\end{align}
the quintic develops 16 conifold singularities. The triangulation stays unique and the K\"ahler cone simplicial, with
\begin{align}
\label{eq:KCQuniticConifold}
J=t_1 J_1+t_2' J_2'\; ,
\end{align}
where $J_1=D_5$ and $J_2'=D_5-D'_6$. Note that now $t_1\to i\infty$ leads to a IV$_2$ rather than a IV$_1$ degeneration. Moreover, the limit $t_2'\to i\infty$ gives rise to a II$_{1}$ degeneration.  The combined limit $t_1,t_2'\to i\infty$ gives rise to a IV$_{2}$ degeneration.

It is also possible to add both vertices $v_6$ and $v_6^\prime$ simultaneously, in which case the CY will have a non-generic dP$_6$ singularity and 12 conifold points. The triangulation stays unique and the K\"ahler cone simplicial with
\begin{align}
\label{eq:KCQuniticdP6Conifold}
J=t_1 J_1+t_2 J_2+t_3 J_3\;,
\end{align}
where $J_1=D_5$, $J_2=D_5-D_6$, and $J_3=D_5-D_6-D'_6$. As in the previous case~\eqref{eq:KCQuniticConifold}, $t_1\to i\infty$ leads to a IV$_2$ degeneration. Moreover, $t_2\to i\infty$ leads to a IV$_2$ as in the first blowup case~\eqref{eq:KCQuniticdP6}. The limit $t_3\to i\infty$ leads to a II$_2$ degeneration; in~\eqref{eq:KCQuniticConifold}, the limit $t_2'\to i\infty$ gave rise to a II$_1$ limit. However, the divisor $J_3=J_2'-D_6$ and the topology changed such that there are 12 instead of 16 conifold singularities. Combining limits $t_i,t_j\to i\infty$ and $t_1,t_2,t_3\to i\infty$ gives rise to IV$_3$ degenerations. 

The enhancement diagrams corresponding to the enhancement chains are given in Figure~\ref{fig:HasseDiagramsQuinticChain}. None of the four compactification spaces discussed here is elliptically fibered, but both the conifold transition and the combined transition of the quintic have a K3 fibration. This can be seen from the presence of a type II but the absence of a type III vertex, as we will discuss in~Section \ref{sec:DiagramsFibrations}.

\subsection{Enhancement diagrams for non-simplicial K\"ahler cones}
\label{sec:NonSimplicaialKCEnhancementDiagrams}
To address all regions in a non-simplicial K\"ahler cone, one can subdivide this non-simplicial cone into simplicial $h^{1,1}$-dimensional subcones. This means we consider all subsets of $h^{1,1}$ linearly independent generators $\omega_{i_1},\ldots, \omega_{i_{h^{1,1}}}$ out of the generators $\omega_i$ that span the non-simplicial cone. For each of these subcones one can then determine an enhancement pattern by following the strategy for simplicial K\"ahler cones outlined above. By keeping track of how generators are shared among subcones, we can thereafter patch all these different enhancement patterns together, resulting in an enhancement pattern for the whole non-simplicial K\"ahler cone, for example
\begin{align}\label{eq:patternnonsimplexample}
&\big( (), \text{I}_0 \big), \big((\omega_1) , \text{IV}_1 \big), \big( (\omega_2),  \text{IV}_2 \big), \big( (\omega_3),\text{IV}_2\big) ,\big( (\omega_4), \text{IV}_3 \big) ,\nonumber\\
&\big( (\omega_1,\omega_2) , \text{IV}_2 \big), \big( (\omega_1,\omega_3) , \text{IV}_2 \big), \big( (\omega_1,\omega_4) , \text{IV}_3 \big), \big((\omega_2,\omega_3) , \text{IV}_3 \big), \big((\omega_2,\omega_4) , \text{IV}_3 \big) , \big( (\omega_3,\omega_4) , \text{IV}_3 \big) , \\
&\big((\omega_1,\omega_2,\omega_3) , \text{IV}_3 \big), \big((\omega_1,\omega_2,\omega_4) , \text{IV}_3 \big), \big((\omega_1,\omega_3,\omega_4) , \text{IV}_3 \big) ,\big( (\omega_2,\omega_3,\omega_4) , \text{IV}_3 \big) \; ,\nonumber
\end{align}
where each set of generators in the last row indicates a different three-dimensional simplicial subcone. We note that the degeneration type for every $h^{1,1}$-dimensional simplicial subcone, when all its moduli are sent to their limit, will always be IV$_{h^{1,1}}$, just like for a simplicial K\"ahler cone.

\begin{figure}[t]
\centering
\includegraphics[width=6cm]{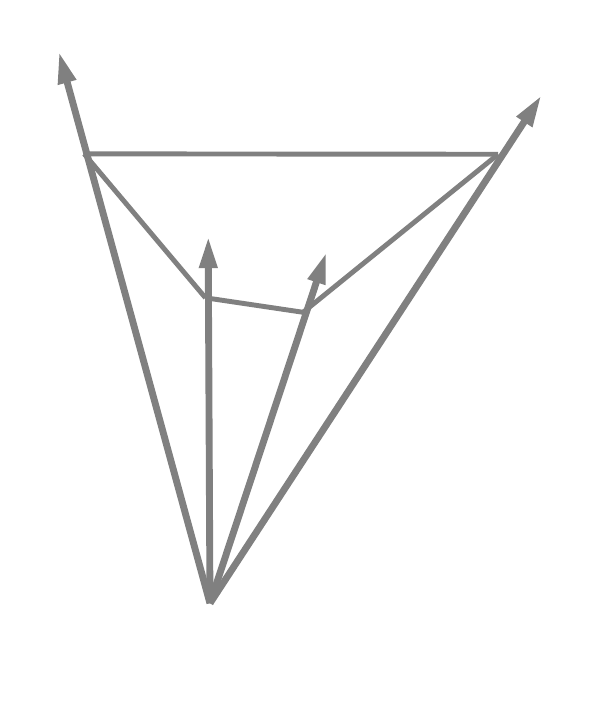}
\begin{picture}(0,0)
\put(-120,140){$\omega_1$}%1
\put(-87,137){$\omega_3$}%3
\put(-164,194){$\omega_2$}%2
\put(-18,180){$\omega_4$}%4
\end{picture}\hspace{10mm}
\includegraphics[width=6cm]{nonsimplicialcone.pdf} 
\begin{picture}(0,0)
\put(-127,96){\tiny IV$_1$}%1
\put(-140,76){\tiny IV$_2$}%2
\put(-100,136){\tiny IV$_3$}%3
\put(-105,108){\tiny IV$_2$}%4
\put(-88,96){\tiny IV$_2$}%1,2
\put(-135,125){\tiny IV$_2$}%1,3
\put(-67,96){\tiny IV$_3$}%2,4
\put(-69,125){\tiny IV$_3$}%3,4
\put(-100,166){\tiny IV$_3$}%interior
\end{picture}
\vspace{-12mm}
\caption{Example of a non-simplicial K\"ahler cone with generators $\omega_i,~i=1,2,3,4$. Each face is labeled with its corresponding enhancement types. Note that all faces in the interior of the cone, such as $(\omega_1,\omega_4)$, have not been labeled explicitly, since the limit type of a generic point in the interior is always IV$_{h^{1,1}}$.}\label{fig:nonsimplicialexample}
\end{figure}

To deal with the non-simplicial cone in this manner, it is crucial that the degeneration type of a limit does not depend on the choice of simplicial subcone in which this limit is considered. That is, for a limit of the generators $\omega_i$ with $i\in\cI =(i_1,\ldots, i_n )$, it does not matter which additional (independent) generators $\omega_{i_{n+1}},\ldots, \omega_{i_{h^{1,1}}}$ we choose to compute its limit type. This follows from the fact that the degeneration type can be determined via the three quantities $\rk \cK^{(\cI)}, \rk \cK^{(\cI)}_I,\rk \cK^{(\cI)}_{IJ}$ (see Table \ref{Type_Table2}), and these are invariant under the choice of basis for $H^2(Y_3)$. Altogether, this allows us to label the faces of the non-simplicial K\"ahler by degeneration types, as exemplified in Figure \ref{fig:nonsimplicialexample}, in a fashion similar to the simplicial cone.

\begin{figure}[t]
\centering
\subfloat[Hasse diagram.]{\label{fig:hassediagramnonsimplexampleA}\includegraphics[width=8cm]{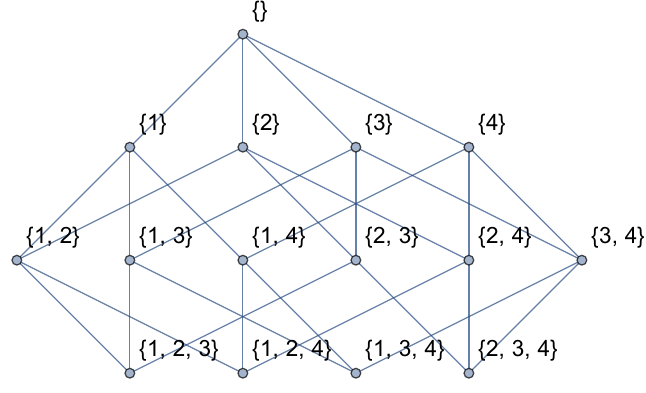}}
\hspace{1cm}
\subfloat[Enhancement diagram.]{\label{fig:hassediagramnonsimplexampleB}\includegraphics[width=8cm]{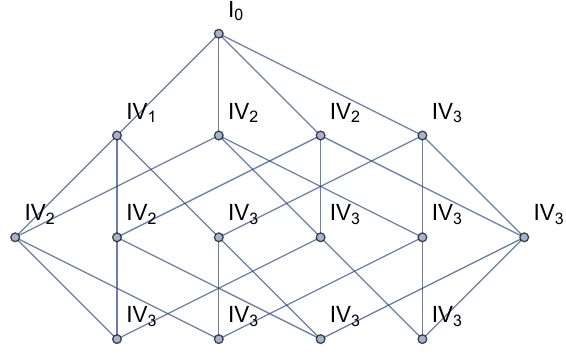}}
\caption{Hasse and enhancement diagrams for Calabi-Yau threefolds with non-simplicial K\"ahler cones. On the left, we give the set of generators and on the right the enhancement diagram for the enhancement pattern of \eqref{eq:patternnonsimplexample}. This diagram corresponds (for instance) to polytope 43 in the Kreuzer-Skarke list (its fine regular star triangulation is unique).}\label{fig:hassediagramnonsimplexample}
\end{figure}

Next we can repackage this enhancement pattern into an enhancement diagram, analogous to the simplicial K\"ahler cone. Namely, we can first consider the Hasse diagram associated with (the power set of) the total set of generators of the non-simplicial cone, keeping only vertices corresponding to independent sets of generators, which results in a diagram as displayed in Figure \ref{fig:hassediagramnonsimplexample}. Then, one can label all the vertices with the corresponding degeneration type as given by the corresponding enhancement pattern, which results in an enhancement diagram such as in Figure \ref{fig:hassediagramnonsimplexample}. Note that each of the vertices in the last row of these diagrams will correspond to one of the simplicial $h^{1,1}$-dimensional subcones, and by considering all enhancement chains that end at such a vertex, i.e.~all downward-moving paths in the diagram, one can recover the enhancement diagram for each of these subcones as a subgraph.

\subsection{On symmetries of the enhancement diagrams}\label{sec:symmetries}
\begin{figure}[t]
\centering
\subfloat[Example 1.]{\label{fig:hassediagramsymmetriesA}\includegraphics[width=8cm]{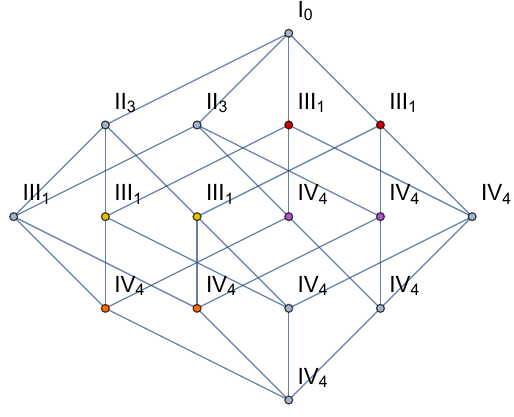}}\qquad
\subfloat[Example 2.]{\label{fig:hassediagramsymmetriesB}\includegraphics[width=8cm]{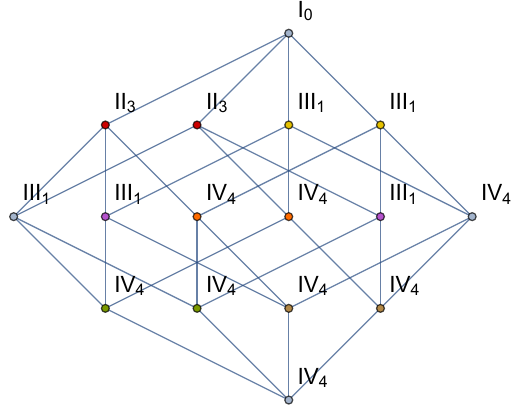}}
\caption{Two enhancement diagrams for the enhancement patterns in \eqref{eq:symmetrypatterns}, where indistinguishable sets of vertices, i.e.~vertices with the same subgraphs in the sense of Figure \ref{fig:subgraphhighlight}, have been highlighted in the same color. The first diagram is found for CICY 7860, and the second diagram for instance for polytope 288 in the Kreuzer-Skarke list (its fine regular star triangulation is unique) and for CICY 7864.}\label{fig:hassediagramsymmetries}
\end{figure}

Now that we have established how to construct enhancement diagrams when given the intersection numbers of a Calabi-Yau threefold, we take a closer look at these graphs, and in particular their symmetries. To illustrate the first part of this discussion, we use the two enhancement patterns
\begin{align}\label{eq:symmetrypatterns}
&\big( (), \text{I}_0 \big), \big((\omega_1) , \text{II}_3 \big), \big( (\omega_2),  \text{II}_3 \big), \big( (\omega_3),\text{III}_1\big) ,\big( (\omega_4), \text{III}_1 \big) ,\nonumber\\
&\big( (\omega_1,\omega_2) , \text{III}_1 \big), \big( (\omega_1,\omega_3) , \text{III}_1 \big), \big( (\omega_1,\omega_4) , \text{III}_1 \big), \big((\omega_2,\omega_3) , \text{IV}_4 \big), \big((\omega_2,\omega_4) , \text{IV}_4 \big) , \big( (\omega_3,\omega_4) , \text{IV}_4 \big), \nonumber \\ 
&\big((\omega_1,\omega_2,\omega_3) , \text{IV}_4 \big), \big((\omega_1,\omega_2,\omega_4) , \text{IV}_4 \big), \big((\omega_1,\omega_3,\omega_4) , \text{IV}_4 \big) ,\big( (\omega_2,\omega_3,\omega_4) , \text{IV}_4 \big) \; ,\nonumber \\
&~\\ 
&\big( (), \text{I}_0 \big), \big((\omega_1) , \text{II}_3 \big), \big( (\omega_2),  \text{II}_3 \big), \big( (\omega_3),\text{III}_1\big) ,\big( (\omega_4), \text{III}_1 \big) ,\nonumber\\
&\big( (\omega_1,\omega_2) , \text{III}_1 \big), \big( (\omega_1,\omega_3) , \text{III}_1 \big), \big( (\omega_1,\omega_4) , \text{IV}_4 \big), \big((\omega_2,\omega_3) , \text{IV}_4 \big), \big((\omega_2,\omega_4) , \text{III}_1 \big) , \big( (\omega_3,\omega_4) , \text{IV}_4 \big), \nonumber \\ 
&\big((\omega_1,\omega_2,\omega_3) , \text{IV}_4 \big), \big((\omega_1,\omega_2,\omega_4) , \text{IV}_4 \big), \big((\omega_1,\omega_3,\omega_4) , \text{IV}_4 \big) ,\big( (\omega_2,\omega_3,\omega_4) , \text{IV}_4 \big) \; ,\nonumber
\end{align}
whose enhancement diagrams are given in the first and second graph of Figure \ref{fig:hassediagramsymmetries}, respectively.

\begin{figure}[t]
\centering
\includegraphics[width=7.5cm]{symmetrydiagram2.pdf}
\begin{minipage}[b]{0.5\textwidth}
\includegraphics[width=7.5cm]{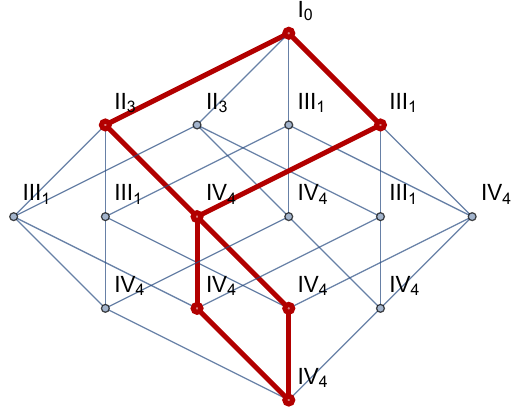} 
\end{minipage}\hspace*{0.2cm}
\begin{minipage}[b]{0.5\textwidth}
\includegraphics[width=7.5cm]{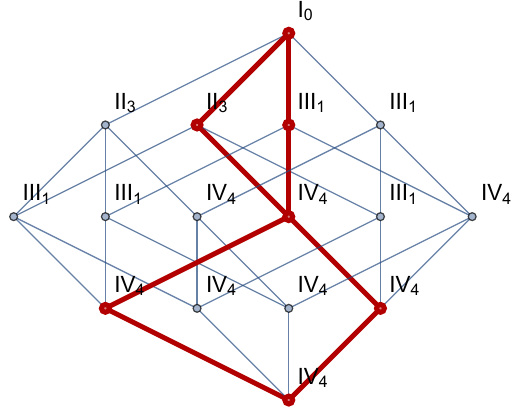} 
\end{minipage}
\caption{Enhancement diagrams (both are the diagram in Figure \ref{fig:hassediagramsymmetriesB}) where the two subgraphs, obtained by considering all enhancement chains through the two indistinguishable Type IV$_4$ vertices in the second row, have been highlighted. Indeed, the two subgraphs are identical.}\label{fig:subgraphhighlight}
\end{figure}

The two enhancement diagrams look very similar at first sight. In order to decide whether they actually are identical, we first perform some simple comparisons. For instance, we find the same sets of degeneration types at each row of the graph. Furthermore, if one reads off all $4!=24$ enhancement chains for each of these diagrams, one obtains two identical sets of enhancement chains. This raises the question whether these two graphs are actually the same or not, and the difference between them boils down to how these chains overlap with each other. To be more precise, one can consider all enhancement chains that share a given vertex, which together form a subgraph, as depicted in Figure \ref{fig:subgraphhighlight}. We notice that certain vertices have the same subgraphs, which means that these vertices are \textit{indistinguishable} and thus give rise to a symmetry of the graph. This proves to be an interesting way of telling two graphs apart. In the example in Figure~\ref{fig:hassediagramsymmetries}, different sets of vertices turn out to be indistinguishable. 

\begin{figure}[t]
\centering
\vspace*{-.2cm}
\includegraphics[width=9cm]{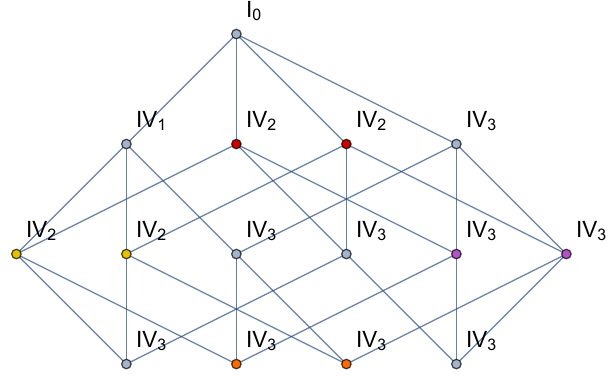}
\vspace*{-.2cm}
\caption{Enhancement diagram for enhancement pattern \eqref{eq:patternnonsimplexample} for a non-simplicial K\"ahler cone (cf.~Figure \ref{fig:hassediagramnonsimplexample}), where all sets of indistinguishable vertices have been highlighted. Note that two vertices at the last row, each representing a simplicial $h^{1,1}$-dimensional subcone, are indistinguishable.}\label{fig:symmetrynonsimpl}
\end{figure}

If vertices at the last row are indistinguishable for non-simplicial K\"ahler cones, this indicates that the $h^{1,1}$-dimensional simplicial subcones associated with these vertices have the same enhancement pattern, which thus results in a notion of indistinguishable $h^{1,1}$-dimensional simplicial subcones, see for instance Figure \ref{fig:symmetrynonsimpl}.

 \begin{figure}[t]
\vspace*{-.2cm}
\centering
\begin{minipage}[b]{0.45\textwidth}
\includegraphics[width=7.5cm]{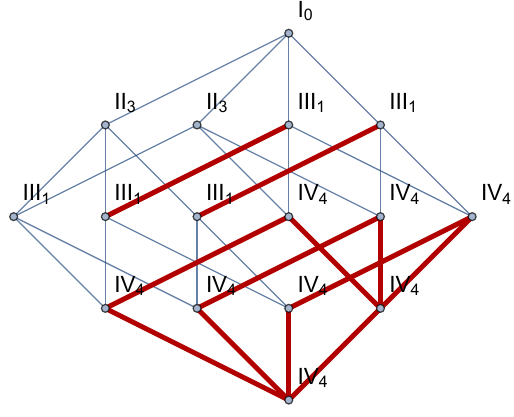} 
\end{minipage}\hspace*{0.2cm}
\begin{minipage}[b]{0.45\textwidth}
\includegraphics[width=7.5cm]{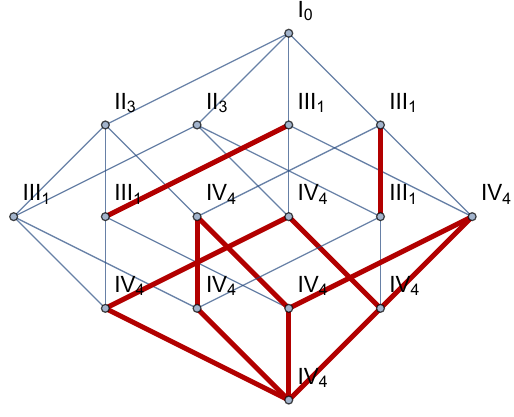} 
\end{minipage}
\vspace*{-0.2cm}
\caption{Enhancement diagrams where the trivial enhancement steps have been highlighted.}\label{fig:trivialsteps}
\end{figure}

\begin{figure}[t]
\vspace*{-.2cm}
\centering
\subfloat[Enhancement diagram.]{\label{fig:diagramreducedA}\hspace{1cm}\includegraphics[height=5cm]{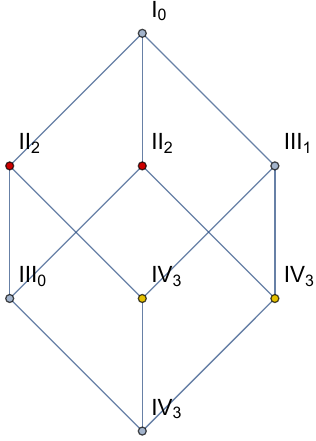}\hspace{1cm}}
\hspace{2cm}
\subfloat[Reduced enhancement diagram.]{\label{fig:diagramreducedB}\hspace{2cm}\includegraphics[height=5cm]{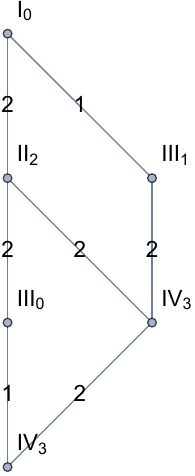}\hspace{2cm}}
\caption{Reduction of an enhancement diagram via its symmetries, where we stack sets of indistinguishable vertices on top of each other. Note that each edge has now been assigned a label, which indicates its multiplicity. This diagram occurs for polytope 119 in the Kreuzer Skarke list (the polytope has a unique fine regular star triangulation) and for CICY 7880.}\label{fig:diagramreduced}
\end{figure}

We can use these symmetries to stack indistinguishable vertices on top of each other. To illustrate this procedure, consider the enhancement pattern
\begin{align}
\label{eq:symmetrypatternextra}
&\big( (), \text{I}_0 \big), \big( (\omega_1),  \text{II}_2 \big), \big( (\omega_2),   \text{II}_2 \big), \big( (\omega_3), \text{III}_1 \big), \nonumber \\
&\big( (\omega_1,\omega_2),\text{III}_0 \big) , \big( (\omega_1,\omega_3),\text{IV}_3 \big) , \big( (\omega_2,\omega_3),  \text{IV}_3 \big) , \big( (\omega_1,\omega_2,\omega_3), \text{IV}_3\big) \; .
\end{align}
Its associated (standard) enhancement diagram is given in Figure~\ref{fig:diagramreducedA}, and the reduced enhancement diagram in Figure~\ref{fig:diagramreducedB}. Note that stacking vertices also results in stacked edges. We therefore label each edge by its multiplicity. One can then deduce whether a vertex in a symmetry-reduced graph represents multiple vertices in the (standard) non-reduced graph via the number of incoming edges $e$ at this vertex from above. For a single vertex located at the $n^\text{th}$ row, there should be $n$ such edges, since a set of $n$ generators can be split up into $n$ different sets consisting of $n-1$ generators, so the multiplicity $m$ of a vertex is $m = e / n$. %To invert the stacking procedure, one must then re-attach all edges accordingly after unstacking the vertices. 
The benefit of this procedure is that it reduces the complexity of the graphs by reducing the number of its vertices. The number of vertices in a graph is $h^{1,1}!$ for simplicial K\"ahler cones, and even larger for a non-simplicial K\"ahler cones. Thus, reducing the number of vertices improves readability of the graphs for large values of $h^{1,1}$. %For instance for geometries with only $\mathrm{IV}_{h^{1,1}}$ limits (and $\mathrm{I}_0$), it would lower the number of vertices to $h^{1,1}+1$. 

Another observation we want to point out is that the degeneration type does not necessarily increase when additional moduli $t^I$ are sent to their limits. We call such cases \textit{trivial enhancements}. Note that the two diagrams of Figure \ref{fig:hassediagramsymmetries} have the same set of trivial enhancements, as highlighted in Figure \ref{fig:trivialsteps}, since these graphs have the same set of enhancement chains. In the spirit of lowering the number of vertices in a graph, one could then stack vertices connected by a trivial enhancement on top of each other and indicate trivial enhancement steps via loops at the stacked vertices.

%%%%%%%%%%%%%%%%%%%%%%%%%%%%%%%%%%%%%%%%%%%%%%%%%%%%%%%%%%%%%%%%%%%%%%%%%%%%
\subsection{Construction of enhancement diagrams via recursion}
\label{sec:recursion}
The degeneration type follows a recursion formula (cf.\ equation \eqref{eq:recursion} below), which determines the index (I, II, III, or IV) of the degeneration type, but not its subindex. The recursion formula can be applied to any vertex below the third row of the enhancement diagram and only requires the index of the degeneration type of the vertices which come before it in its enhancement chains. This means that all indices in the enhancement pattern are fixed by the first three rows. In particular, this fixes the numbers of fibrations, which are determined by counting the numbers of II and III vertices, cf.~Section \ref{sec:DiagramsFibrations}.

Let us consider a vertex $\cI_n=(i_1,\ldots, i_n)$ positioned below the third row ($n> 3$). Recall from Table \ref{Type_Table2} that the index of this vertex is determined via $\rk \cK^{(\cI_n)}$ and $\rk \cK^{(\cI_n)}_I$. By expanding these quantities in sums over triples of generators via \eqref{notation}, and by using that the intersection numbers are non-negative in the K\"ahler cone basis, we obtain
\begin{equation}
\begin{aligned}
\rk \cK^{(\cI_n)} &= \max_{\cJ_{3}\subset \cI_n} \rk \cK^{(\cJ_{3})} \; ,\\
\rk \cK^{(\cI_n)}_{ I} &= \max_{\cJ_{3}\subset \cI_n} \rk \cK^{(\cJ_{3})}_{ I}\; ,\\
\end{aligned}
\end{equation}
where $\cJ_{3}=(j_1,j_2,j_3)$ indicate the triples of generators. This implies that the index of the degeneration type of this vertex is given by
\begin{equation}\label{eq:index}
\text{Type $\hat{\text{A}}$}_{(\cI_n)} = \max_{\cJ_{3} \subset \cI_n}\ \text{Type $\hat{\text{A}}$}_{(\cJ_{3})}\; ,
\end{equation} 
where $\hat{\text{A}}_{(\cI_n)}$, $\hat{\text{A}}_{(\cJ_{3})}$ denote the indices of the degeneration types (without subindex) for limits $\cI_n,\cJ_3$. In other words, we find that the index of a vertex is given by the highest index among all the vertices of the third row it is connected to. Equivalently, this means that the index of any vertex below the third row is given by the largest index occurring in any of its enhancement chains. Using~\eqref{eq:index}, the indices in the first three rows of the enhancement diagram thus allow us to determine the indices of all vertices in the enhancement diagram. 

\begin{figure}[t]
\vspace*{-.2cm}
\centering
\begin{minipage}[b]{0.45\textwidth}
\includegraphics[width=8cm]{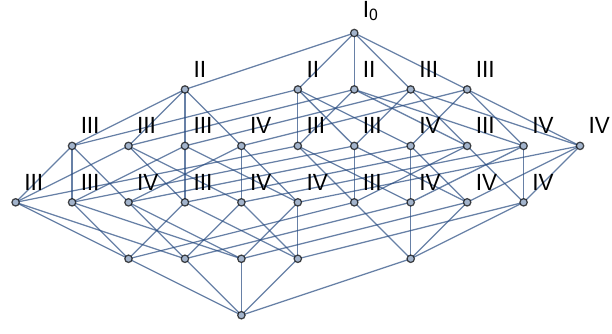} 
\end{minipage}\hspace*{1.5cm}
\begin{minipage}[b]{0.45\textwidth}
\centering
\includegraphics[width=8cm]{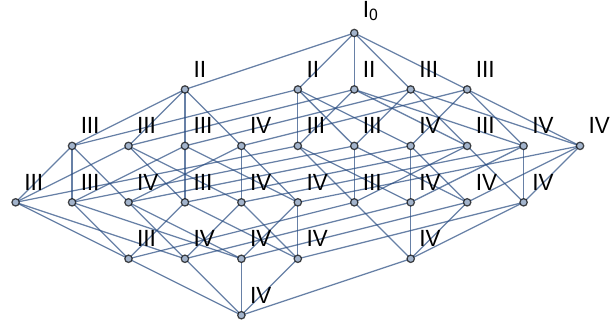} 
\end{minipage}\hspace*{0.2cm}
\vspace*{-0.2cm}
\begin{picture}(0,0)
\put(225,55){\Large $\implies$}
\end{picture}
\caption{Application of \eqref{eq:index} for an enhancement diagram with $h^{1,1}=5$. The index type of all vertices below the third row are determined via the index types at the third row. This enhancement diagram was obtained for CICY 7800.}\label{fig:reconstruct}
\end{figure}
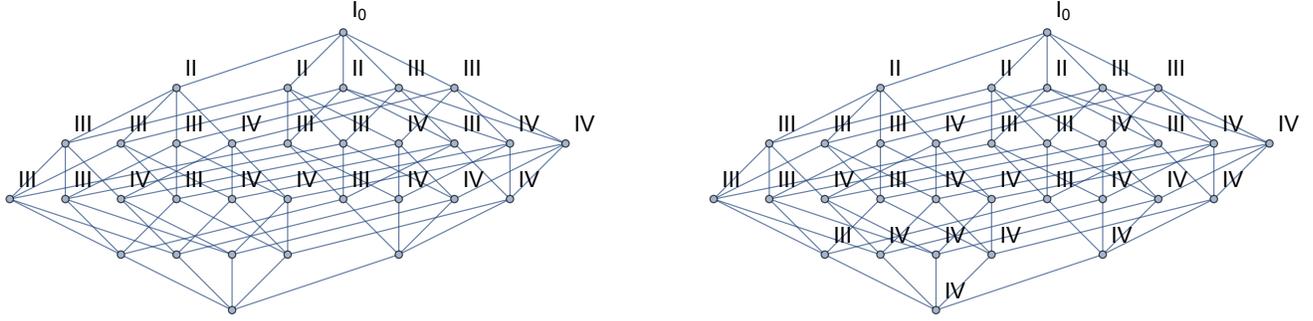

Note that this tells us that there must be a Type IV vertex in the third row of the enhancement diagram, since otherwise we can never obtain the maximal degeneration type IV$_{h^{1,1}}$, which always occurs when all K\"ahler moduli are sent to their limit. Furthermore, since the degeneration type cannot decrease when additional K\"ahler moduli are sent to a limit, we find that $(h^{1,1}-3)!-2$ additional vertices are fixed to be of Type IV for a simplicial K\"ahler cone, and even more for a non-simplicial one. These vertices are connected to this IV vertex at the third row via enhancement chains. An example for how \eqref{eq:index} can be used to determine the enhancement diagram from the third row is given in Figure \ref{fig:reconstruct}. Alternatively this formula can also be recast into the form of a recursion formula
\begin{equation}
\label{eq:recursion}
\text{Type $\hat{\text{A}}$}_{(\cI_{n})} = \max_{\cJ_{n-1} \subset \cI_{n}}\ \text{Type $\hat{\text{A}}$}_{(\cJ_{n-1})}\; ,
\end{equation}
where $\cJ_{n-1}$ are vertices at row $n-1$ connected to vertex $\cI_{n}$ at the $n^\text{th}$ row for $n> 3$. This allows us to determine the indices of the vertices of rows 4 up to $h^{1,1}$ of the enhancement diagram  by iteration.

\section{Enhancement diagrams and fibrations}\label{sec:DiagramsFibrations}
In this section we study how fibrations of CYs are encoded in the enhancement diagrams. The types of possible fibrations of a CY threefold and criteria for when they occur have been classified by Oguiso~\cite{Oguiso:1993aaa}. Which case occurs depends only on the dimension of the base and the value of the integrated second Chern class. One-dimensional fibers can only be elliptic curves, and two-dimensional fibers can be either K3 surfaces or Abelian surfaces (i.e.\ $T^4$).

We use two types of CY constructions: CYs that are constructed as hypersurfaces in a toric ambient space $\mathcal{A}$ (as classified by Kreuzer and Skarke~\cite{Kreuzer:2000xy}), and CYs that are constructed as complete intersections in an ambient space $\mathcal{A}$ that is a product of projective spaces (as classified by Candelas et.al.~\cite{Candelas:1987kf}).

\subsection{Kreuzer-Skarke and CICYs}

The Kreuzer-Skarke CYs are given by the anti-canonical hypersurface in a four-dimensional toric ambient space. The achievement of Kreuzer and Skarke was to classify all reflexive polyhedra corresponding to the ambient spaces, of which there are almost half a billion inequivalent ones. A 4D reflexive polyhedra is specified (up to toric morphisms, which do not change the CY) by its vertices $v_i$, which are vectors in $\mathbbm{Z}^4$.

We will restrict our focus to those with Picard number smaller than 6 and which are favorable. In order to reference them in the text, we just label them with a running ID, starting from 1, using the ordering of~\cite{Kreuzer:2000xy}. By favorable, we mean the following: A part of $H_{1,1}(X,\mathbbm{Z})$ can always be obtained by pulling back elements from $H_{1,1}(\mathcal{A},\mathbbm{Z})$. In favorable cases, the entire second cohomology of the CY $X$ descends from the ambient space $\mathcal{A}$. In non-favorable cases, there can be new, non-toric divisors on the CY that do not descend directly from the ambient space. These cases can be easily identified using Batyrev's construction and the resulting formulas for the Hodge number of CYs given by reflexive polyhedra~\cite{Batyrev:1994hm}. 

While the reflexive polytopes of~\cite{Kreuzer:2000xy} fix the ambient space and a generic section of the anticanonical bundle, we still need to triangulate the polyhedron. This fixes the intersection ring (or, equivalently, the Stanley-Reisner ideal) and the integrated second Chern classes and hence uniquely determine the CY according to Wall's theorem~\cite{Wall:1966aaa}. We will use \texttt{SAGE} to construct all fine regular star triangulations (FRST) of each polytope. We furthermore use \texttt{SAGE} to construct the K\"ahler cone of the ambient variety for each of the FRSTs. Note that the K\"ahler cone can be non-simplicial, which means that it has more generators than its dimension, cf.\ Section~\ref{sec:EnhancementDiagrams}.\footnote{We content ourselves with constructing the K\"ahler cone on the ambient space rather than on the CY hypersurface, which would be much more involved.} Both the number of different FRSTs as well as the number of simplicial subcones of the non-simplicial K\"ahler cone give rise to a sizable number of examples, even when restricting to favorable cases with $h^{1,1}\leq5$, cf.\ Table~\ref{table:KS} and Table~\ref{table:KSnon-simpl}.

Candelas et.al.~\cite{Candelas:1987kf} classified all 7890 inequivalent complete intersection CYs (CICYs) in products of projective ambient spaces. We restrict again to favorable configurations and use the ID assigned in the list~\cite{Candelas:1987kf,Anderson:2017aux}. For CICYs, it is often possible to split projective ambient space factors and the corresponding normal bundle into products of smaller ambient spaces while not changing the CY. In this way, a favorable description of almost all CICYs as intersecting hypersurfaces in products of projective spaces can be obtained~\cite{Anderson:2017aux}, which we will use here. Since there is no triangulation ambiguity for CICYs and the K\"ahler cone of the ambient space is simplicial for the favorable CICYs we consider here, the subtleties discussed above for toric constructions do not arise. This means a CICY is completely specified in terms of the dimensions of the projective ambient space factors and the degrees of the normal bundles of hypersurfaces that define the complete intersection. Since the normal bundles need to cancel the anti-canonical class in order to obtain a CY, the sum of the degrees of the normal bundle of each ambient space $\mathbbm{P}^{n_i}$ factor needs to be equal to $n_i+1$. This allows us to specify a CICY by just specifying the degrees of the normal bundles on each $P^{n_i}$ ambient space factor. In general we have an ambient space 
\begin{align}
\mathcal{A}=\mathbbm{P}^{n_1}\times\mathbbm{P}^{n_2}\times\ldots\times\mathbbm{P}^{n_K}\; ,
\end{align}
whose dimension is
\begin{align}
D=\sum_{i=1}^K n_i\; .
\end{align}
Since we are interested in favorable threefolds we get $K=h^{1,1}$ (as every ambient space hyperplane class pulls back to a generator of $H^{1,1}(X,\mathbbm{Z})$ on the CICY $X$) and we need to specify $d=D-3$ normal bundles. This allows us to specify a CICY in terms of a $K\times d$ integer matrix called configuration matrix. The number of CICY geometries is given in Table~\ref{table:CICY}.

Note that, as in the toric case, it is possible that the K\"ahler cone of the CY differs from the K\"ahler cone of the ambient space; it can be larger and/or non-simplicial. These cases are referred to as ``non-K\"ahler favorable'' in~\cite{Anderson:2017aux}. In particular, this happens if part of the CICY defines a del Pezzo surface dP$_i$ (which can have non-simplicial K\"ahler cones), and the CICY can be thought of as an intersection inside this dP$_i$ times some other (projective) ambient space factors. For CICYs that have a favorable description in products of projective ambient spaces, the del Pezzos that can occur are dP$_i$, $i=0,1,2,3$. 

\subsection{Elliptic fibrations}
\label{sec:EllFib}
Let us first illustrate how elliptic (or genus one) fibrations are encoded in the enhancement diagrams. We observe the following correspondence:
\begin{enumerate}
\item The total number of elliptic fibrations is given by the number of Type III$_c$ vertices in the enhancement diagram
\item The number of elliptic fibrations up to birational equivalence of the base is given by the number of connected subgraphs of Type III$_c$ vertices in the enhancement diagram
\end{enumerate}
We explain this in more detail and provide an example in Section~\ref{sec:ExampleEllFib}. Before we discuss examples, we explain how we identify elliptic fibrations.

\subsubsection{Identifying elliptic fibrations}
\label{ssec:EllipticFibrationsKollar}
In order to identify elliptic fibrations of a CY $X$ we can follow different techniques. As was conjectured by Kollar in~\cite{Kollar:2012pv}, $X$ is genus-one fibered (i.e.\ it has a torus fibration which does not necessarily have a section) iff there exists a (1,1)-form $D\in H^2(X,\mathbbm{Q})$ such that 
\begin{align}
\label{eq:Kollar}
D.C\geq0~~~\forall \text{~algebraic curves~} C\subset X\; , \qquad D^3=0\; ,\qquad D^2\neq0 \;.
\end{align}
This was actually proven by Oguiso~\cite{Oguiso:1993aaa} and Wilson~\cite{Wilson:1994aaa} under the additional constraint that 
\begin{align}
D~\text{is effective}\qquad\text{or}\qquad\int_X c_2(X)\cdot D\neq0\;.
\end{align}
The idea behind the criterion is that $D$ is the pullback of an ample divisor in the base. Since it does not have any components in the fiber direction, $D^3=0$. Note that if a divisor $D$ satisfies~\eqref{eq:Kollar}, so does any positive multiple of $D$. In~\cite{Anderson:2017aux}, this redundancy is dealt with by considering two elliptic fibrations with divisors $D$ and $D'$ as equivalent if
\begin{align}
\label{eq:EquivalenceKollar}
D^2\sim D'^2~\text{as curves in}~X\;.
\end{align}
This equivalence can be checked either directly by comparing the corresponding expressions of $D^2$ and $D'^2$ in cohomology modulo the Stanley-Reisner ideal and linear equivalences, or by comparing the triple intersections $D^2.D_i$ with $D'^2.D_i$, where $D_i$, $i=1,\ldots,h^{1,1}(X)$ is a basis of $H^2(X)$. Note that~\eqref{eq:EquivalenceKollar} can be true over generic points, but at special points over the base the fibrations could still be different, which happens for birationally equivalent bases.

For toric varieties, there exists another sufficient (but not necessary) criterion for elliptic fibrations. As was noted in~\cite{Avram:1996pj}, if the polytope defining the toric ambient space contains one of the 16 reflexive polytopes in 2D along a hypersurface through the origin and there exists a compatible FRST, then this ambient space fibration is inherited by the CY and the anticanonical hypersurface of the 2D polytope specifies the fiber of the genus one fibration of the CY. Such ``toric genus one fibrations'' are particularly nice since they allow us to use powerful and well-established techniques to study properties of the fibration (such as its degenerations in various codimensions, the rank of the Mordell-Weil group, $\ldots$). However, we find that most of elliptic fibrations as identified by Kollar's criterion, are not of this type.

For CICYs, there is also an alternative way to identify elliptic fibrations, dubbed obvious elliptic fibrations in~\cite{Anderson:2017aux}. Again, the idea is to identify those equations within the whole set of complete intersection equations that define the fiber. This can be done easily on the level of the configuration matrix. If the matrix can be rearranged to take the form
\begin{align}
\label{eq:OGF}
\left[
\begin{array}{l|cc}
\mathcal{A}_1&0&\mathcal{F}\\
\mathcal{A}_2&\mathcal{B}&\mathcal{T}\\
\end{array}
\right]\;,
\end{align}
the genus one fiber is given by the complete intersection $[\mathcal{A}_1~|~\mathcal{F}]$ and the base is given by the complete intersection $[\mathcal{A}_1~|~\mathcal{B}]$, where $\mathcal{T}$ determines how the base is fibered (if $\mathcal{T}$ a zero matrix, the fibration is trivial, i.e.\ a direct product). Moreover, $\mathcal{B}$ can be the empty set, in which case the base is given by just the ambient space factors $\mathcal{A}_2$.  It was observed in~\cite{Anderson:2017aux} by comparing results using Kollar's criterion~\eqref{eq:Kollar} and identifying fibrations via~\eqref{eq:OGF}, that the two methods agree for K\"ahler favorable CICYs, i.e.\ for CICYs all elliptic fibrations can be written in the form~\eqref{eq:OGF}. For toric methods, the ambient space projection method severely underestimates the number of fibrations as obtained by Kollar.

\medskip

Now we are in a position to explain why a Type III$_c$ vertex in the enhancement diagram indicates an elliptic fibration. As discussed in Section \ref{ssec:EllipticFibrationsKollar}, one can deduce whether a Calabi-Yau threefold is elliptically fibered from its topological data. Namely, there should exist a nef divisor, such that its intersection numbers satisfy \eqref{eq:Kollar}. Expanding this divisor in generators $\cI=(\omega_{i_1}, \ldots, \omega_{i_n})$ (with positive coefficients), we obtain straightforwardly the following two conditions on the intersection numbers
\begin{equation}
 \rk \cK^{\cI} = 0\; , \qquad \rk \cK^{\cI}_I =  1\; ,
\end{equation}
using the short-hand notation for $\cK^{\cI}, \cK^{\cI}_I$ defined in \eqref{notation}. These are exactly the two conditions that indicate a Type $\mathrm{III}_c$ limit for the large volume regime, as follows from Table \ref{Type_Table2}. Thus the presence of a Type III$_c$ limit in the enhancement diagram of a Calabi-Yau threefold directly indicates whether it is elliptically fibered

\subsubsection{Example: Enhancement diagrams of elliptically fibered CYs}
\label{sec:ExampleEllFib}
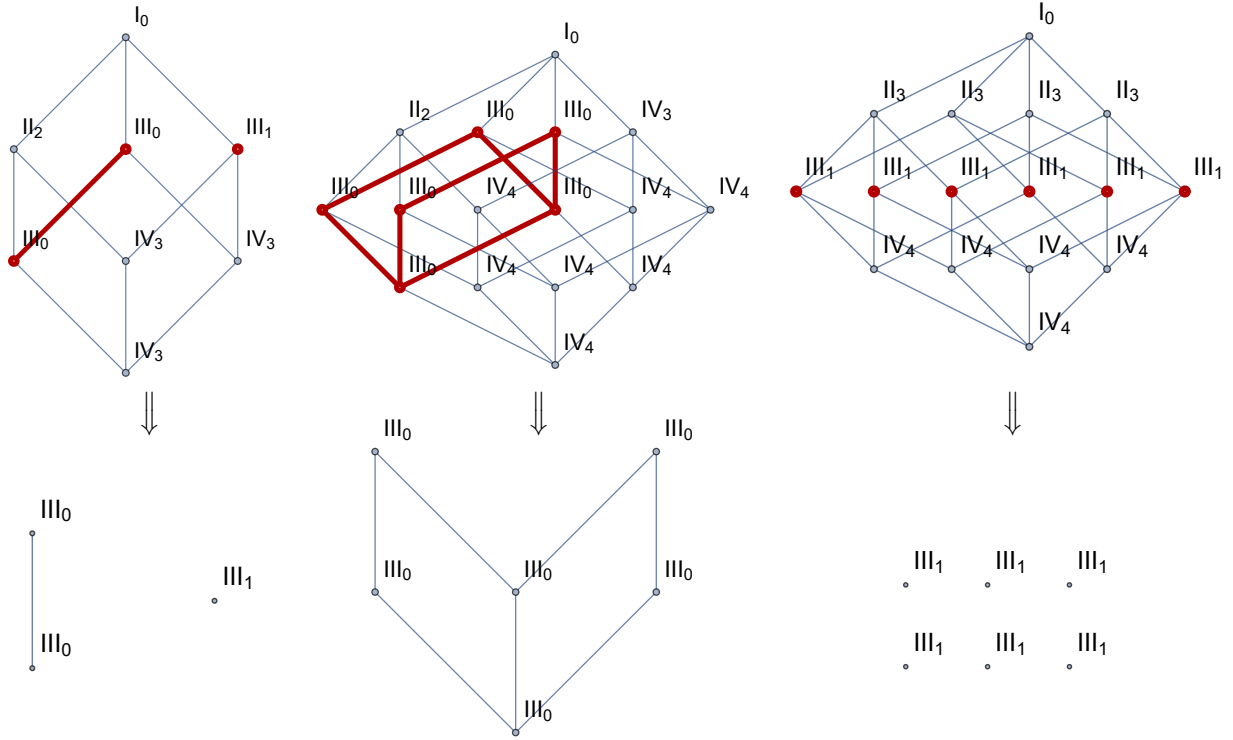
\begin{figure}[t]
\centering
  \begin{minipage}[b]{0.21\textwidth}
\centering
    \includegraphics[width=\textwidth]{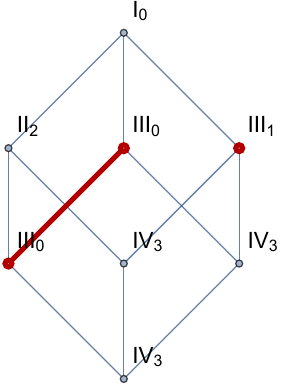}\vspace*{1.5cm} 
    \includegraphics[width=0.85\textwidth]{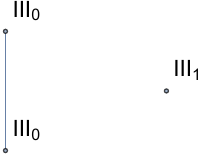}\vspace*{0.9cm} 
    
\begin{picture}(0,0)
\put(0,150){\rotatebox{270}{$\implies$}}
\end{picture}
  \end{minipage}\hspace*{0.3cm}
 \begin{minipage}[b]{0.34\textwidth}
\centering
    \includegraphics[width=\textwidth]{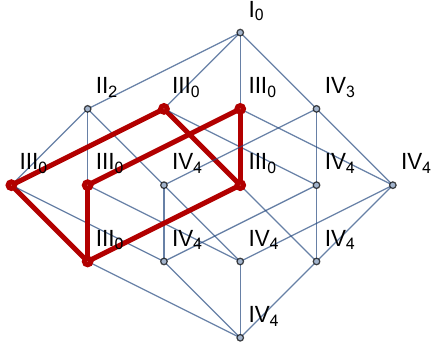}\vspace*{0.5cm}  
    \includegraphics[width=0.75\textwidth]{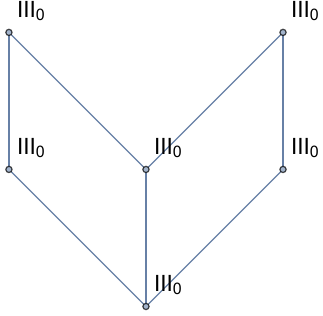} 

\begin{picture}(0,0)
\put(0,150){\rotatebox{270}{$\implies$}}
\end{picture}
  \end{minipage}\hspace*{0.3cm}
 \begin{minipage}[b]{0.34\textwidth}
\centering
    \includegraphics[width=\textwidth]{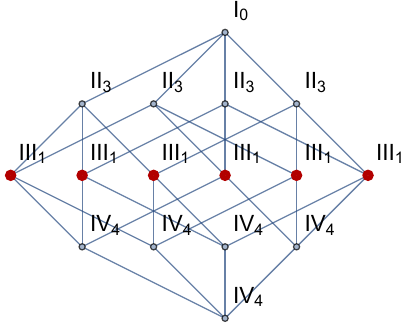}\vspace*{2.5cm}  
    \includegraphics[width=2.8cm]{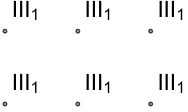}\vspace*{0.9cm}   

\begin{picture}(0,0)
\put(0,150){\rotatebox{270}{$\implies$}}
\end{picture}
  \end{minipage}
\caption{Three example enhancement diagrams to identify elliptic fibrations and elliptic fibrations up to birational equivalence of the base.}\label{fig:graphcountfibrations}
\end{figure}

As an example, we give three different enhancement diagrams in Figure~\ref{fig:graphcountfibrations}. The CYs for which the diagrams were computed are
\begin{enumerate}
\item The first example is obtained from CICY 7875, and from KS polytope 103 or 111. The fine regular star triangulations of these polytopes is unique.
\item The second example is obtained from KS polytope 1460 with Stanley-Reisner ideal \begin{align}\text{SR}=\langle z_0 z_1 ,\, z_0 z_2 ,\, z_3 z_6 ,\, z_2 z_7 ,\, z_2 z_6,\, z_4 z_5 z_7,\, z_1 z_3 z_4 z_5 \rangle\end{align} (the coordinates are ordered in the same way as the vertices are ordered in~\cite{Kreuzer:2000xy}).
\item The third example is obtained from CICY 7862, and from KS polytopes 290 and 647.
\end{enumerate}

For each of the three examples, vertices of Type III limits and the edges between these vertices are highlighted in red in the top figure and drawn separately in the bottom Figure. Bu counting the number of type III vertices, we find that the three example CYs have 3, 6 and 6 distinct elliptic fibrations, respectively. The number of elliptic fibrations with birationally equivalent bases are given by counting the number of connected subgraphs. We find 2, 1, and 6  in the three examples, respectively.

Let us illustrate how to see these fibrations in more detail for the first example in Figure~\ref{fig:graphcountfibrations}. This diagram, corresponding to CICY 7875 in the list of~\cite{Candelas:1987kf,Anderson:2017aux}, has the configuration matrix
\begin{align}
\label{eq:ellFibExampleCICY}
X\sim\left[
\begin{array}{c|cc}
\mathbbm{P}^2&0&3\\
\mathbbm{P}^1&1&1\\
\mathbbm{P}^2&1&2
\end{array}
\right]\; ,
\end{align}
where we have already reordered the rows to make the fibrations more apparent. Using the method outlined above around equation~\eqref{eq:OGF}, we can now identify three elliptic fibrations:

\textbf{Fibration 1.} The elliptic fiber is given by the first row and second column of~\eqref{eq:ellFibExampleCICY}, i.e.\ it is a genus one fibration where the elliptic curve $E$ is given by a cubic in $\mathbbm{P}^2$ and the base $B$ is a complete intersection of bi-degree $(1,1)$ in $\mathbbm{P^1}\times\mathbbm{P}^2$,
\begin{align}
E\sim\left[\begin{array}{c|c}
\mathbbm{P}^2&3
\end{array}
\right]\;,\qquad
B\sim\left[\begin{array}{c|c}
\mathbbm{P}^1&1\\
\mathbbm{P}^2&1
\end{array}
\right]\; .
\end{align}
The base $B$ can be seen to be $\mathbbm{P}^2$ blown up at one point, i.e.\ $B\simeq\text{dP}_1$.

\textbf{Fibration 2.} The elliptic fiber is given by the first two rows and both columns of~\eqref{eq:ellFibExampleCICY}, i.e.\ the elliptic curve $E$ is given by a complete intersection and the base is just the last ambient space $\mathbbm{P}^2$ factor,
\begin{align}
E\sim\left[\begin{array}{c|cc}
\mathbbm{P}^2&0&3\\
\mathbbm{P}^1&1&1
\end{array}
\right]\;,\qquad
B\sim\left[\begin{array}{c}
\mathbbm{P}^2
\end{array}
\right]\;.
\end{align}

\textbf{Fibration 3.} The elliptic fiber is given by the last two rows and both columns of~\eqref{eq:ellFibExampleCICY}, i.e.\ the elliptic curve $E$ is given by a complete intersection and the base is just the first ambient space $\mathbbm{P}^2$ factor,
\begin{align}
E\sim\left[\begin{array}{c|cc}
\mathbbm{P}^1&1&1\\
\mathbbm{P}^2&1&2
\end{array}
\right]\;,\qquad
B\sim\left[\begin{array}{c}
\mathbbm{P}^2
\end{array}
\right]\;.
\end{align}

As we can see, fibration 1 and 2 are related by including the second ambient space factor, i.e.\ the $\mathbbm{P}^1$ factor, in the base or the fiber, respectively. In the former case, the base is dP$_1$ and in the latter, it is a $\mathbbm{P}^2$. These two fibrations correspond to the left and middle Type III$_0$ limits in~\eqref{eq:ellFibExampleCICY}, and the bases are birationally equivalent. Fibration 3 is not related in this way to the other elliptic fibrations and corresponds to the right Type III$_1$ vertex in Figure~\ref{fig:graphcountfibrations}.

\medskip
The CY can also be realized as a hypersurface in a 4D toric ambient space. It has realizations in terms of KS polytope 103 and 111. Using Wall's theorem, the CICY is found to be equivalent to the polytope with ID 111 and vertices
\begin{align}
\label{eq:ellFibExampleToric}
v_1\!=\!\begin{pmatrix}
1\\0\\0\\0
\end{pmatrix}
\,,~
v_2\!=\!\begin{pmatrix}
0\\1\\0\\0
\end{pmatrix}
\,,~
v_3\!=\!\begin{pmatrix}
-1\\-1\\0\\0
\end{pmatrix}
\,,~
v_4\!=\!\begin{pmatrix}
0\\0\\1\\0
\end{pmatrix}
\,,~
v_5\!=\!\begin{pmatrix}
0\\0\\-1\\0
\end{pmatrix}
\,,~
v_6\!=\!\begin{pmatrix}
0\\0\\0\\1
\end{pmatrix}
\,,~
v_7\!=\!\begin{pmatrix}
0\\0\\1\\-1
\end{pmatrix}
\;.
\end{align}
In order to identify elliptic fibrations, we need to find a reflexive 2D subpolytope within the 4D reflexive polytope defined by~\eqref{eq:ellFibExampleToric}. This subpolytope is given at the intersection  of the lattice polytope with two codimension 1 hyperplanes.

\begin{figure}[t]
\centering
\subfloat[$\mathbbm{P}^1$.]{\label{fig:ToricDiagramsP1}\includegraphics[height=0.2\textwidth]{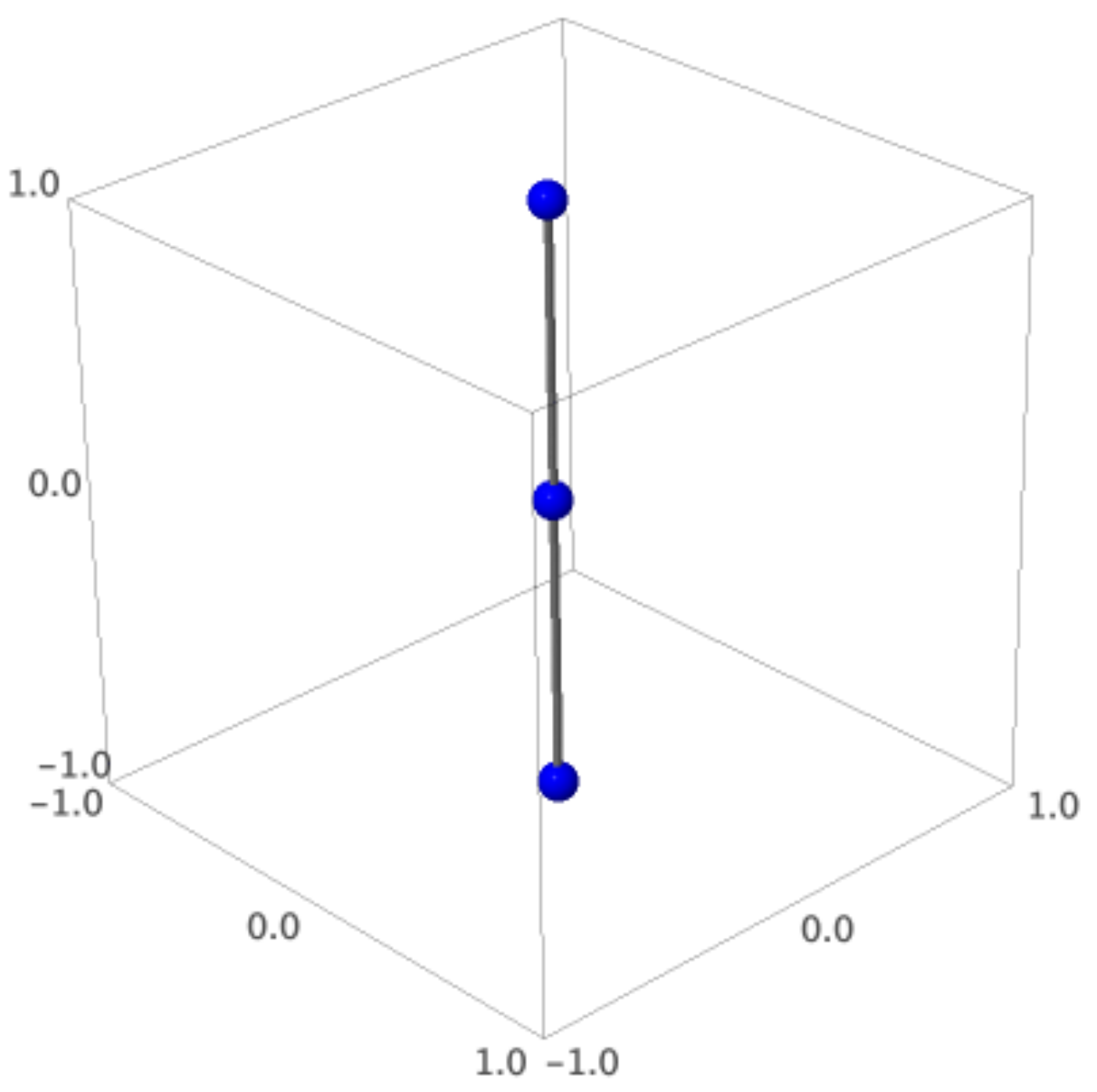}}\quad
\subfloat[$\mathbbm{P}^2$.]{\label{fig:ToricDiagramsP2}\includegraphics[height=0.2\textwidth]{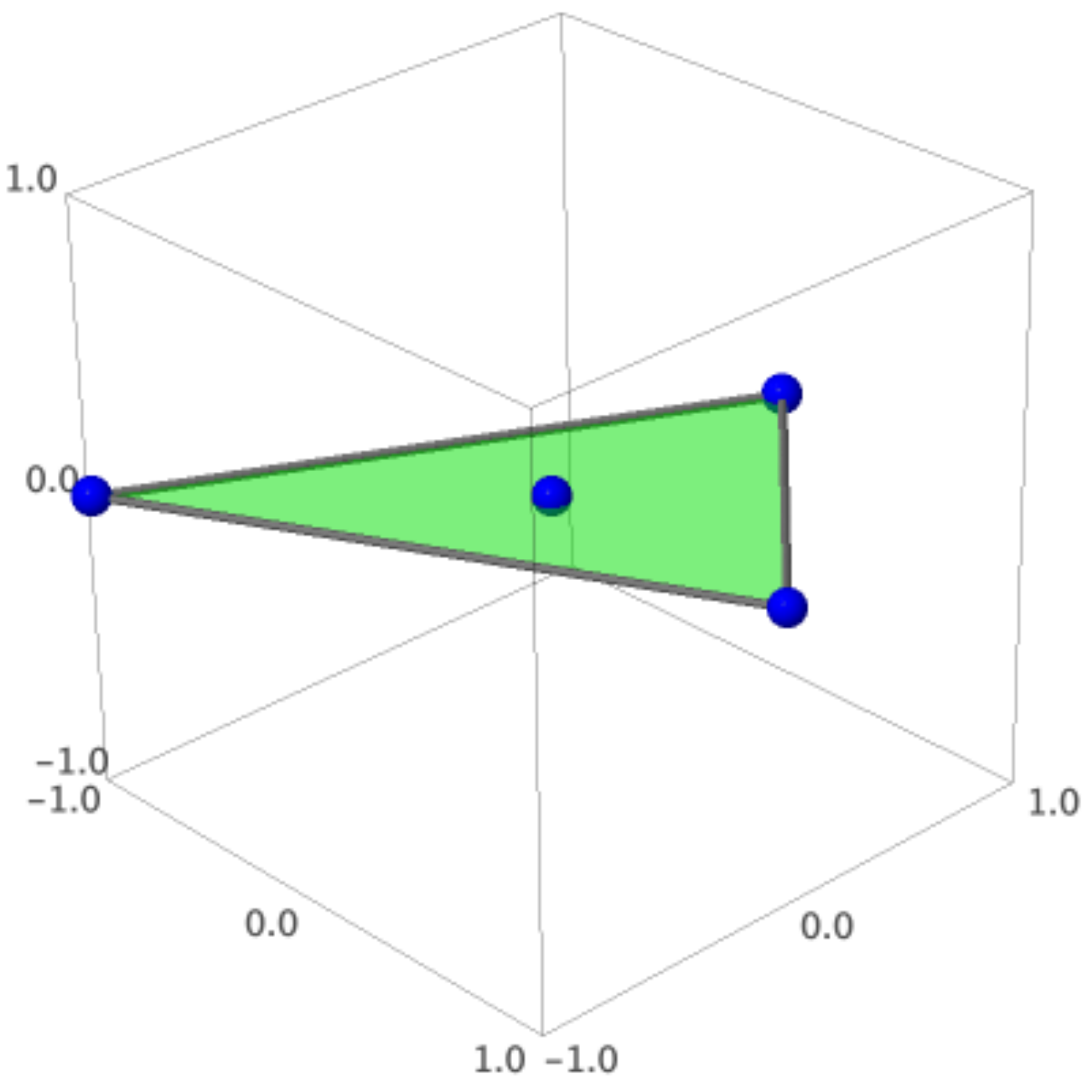}}\quad
\subfloat[dP$_1$.]{\label{fig:ToricDiagramsdP1}\includegraphics[height=0.2\textwidth]{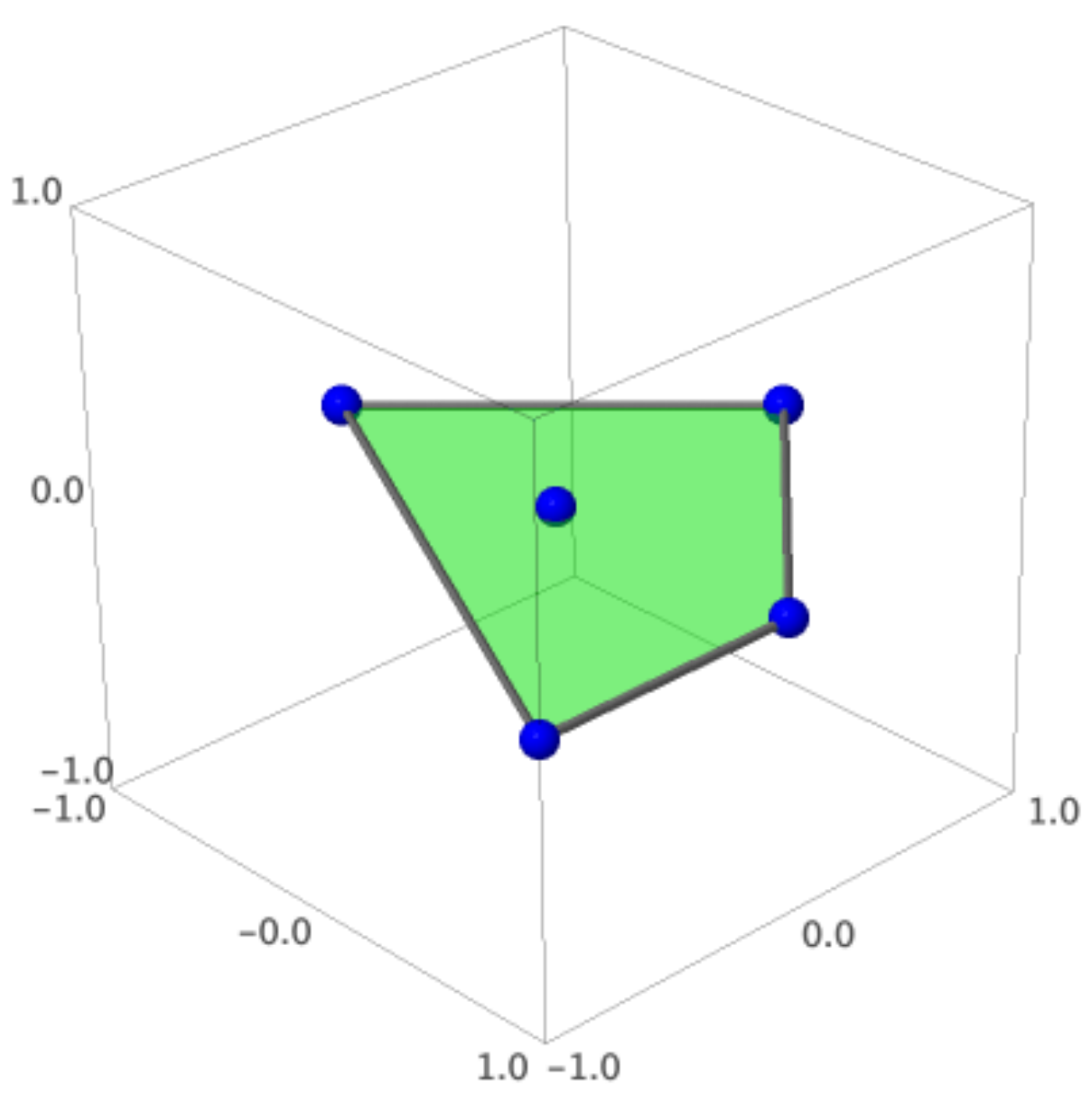}}\quad
\subfloat[K3.]{\label{fig:ToricDiagramsK3}\includegraphics[height=0.2\textwidth]{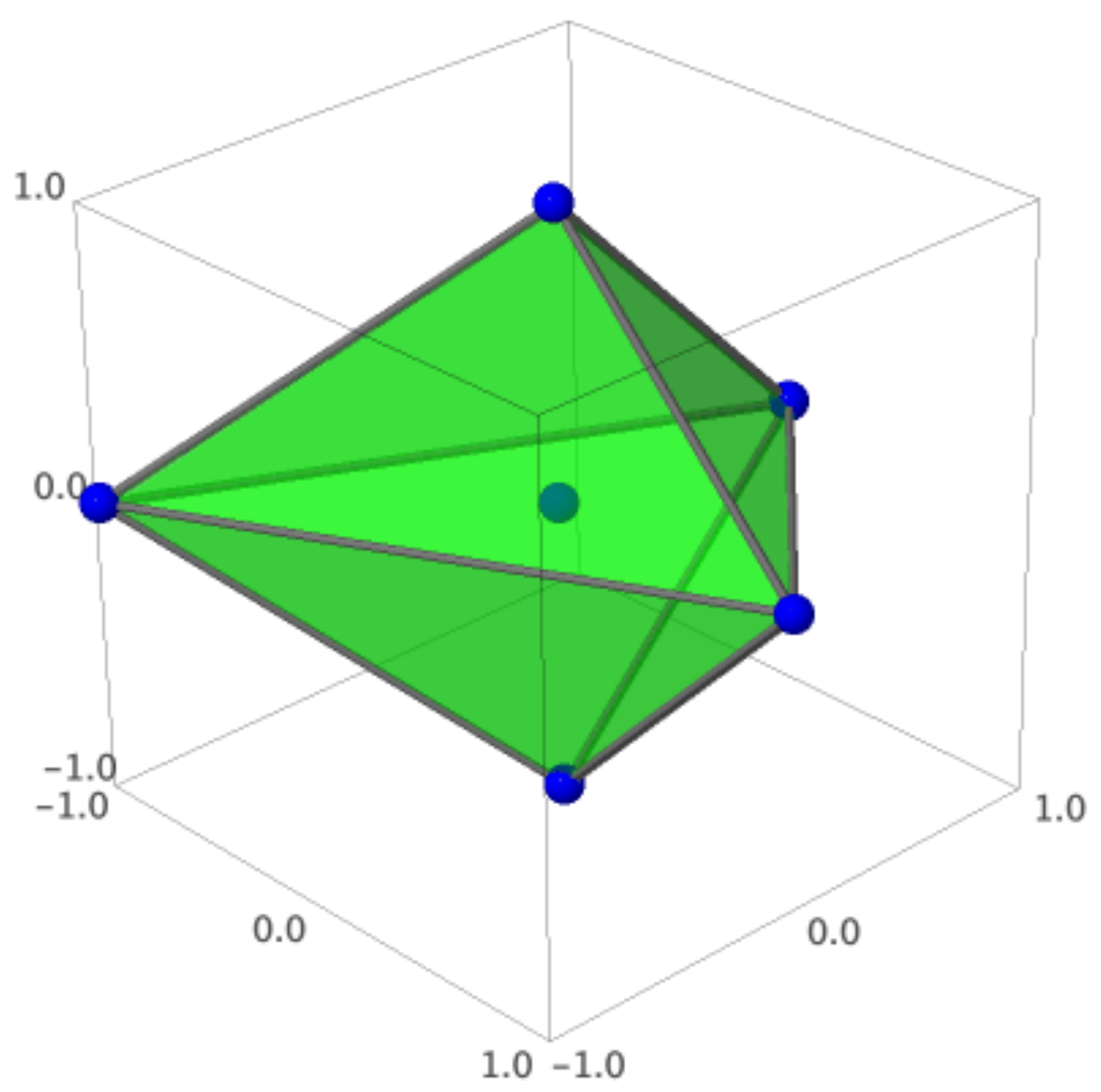}}\,.
\label{fig:ToricDiagrams}
\caption{}
\end{figure}

\textbf{Fibration 1.} We first cut the polytope along the hyperplanes with normal directions $(0,0,1,0)$ and $(0,0,0,1)$, i.e.\ the fiber is given by the first two coordinate entries and the base by the last two entries:
\begin{align}
E&\sim\left\{v_1\!=\!\begin{pmatrix}
1\\0
\end{pmatrix}
\,,~
v_2\!=\!\begin{pmatrix}
0\\1
\end{pmatrix}
\,,~
v_3\!=\!\begin{pmatrix}
-1\\-1
\end{pmatrix}
\,,~
v_4\!=\!\begin{pmatrix}
0\\0
\end{pmatrix}
\,,~
v_5\!=\!\begin{pmatrix}
0\\0
\end{pmatrix}
\,,~
v_6\!=\!\begin{pmatrix}
0\\0
\end{pmatrix}
\,,~
v_7\!=\!\begin{pmatrix}
0\\0
\end{pmatrix}
\right\}
\,,\nonumber\\
B&\sim\left\{v_1\!=\!\begin{pmatrix}
0\\0
\end{pmatrix}
\,,~
v_2\!=\!\begin{pmatrix}
0\\0
\end{pmatrix}
\,,~
v_3\!=\!\begin{pmatrix}
0\\0
\end{pmatrix}
\,,~
v_4\!=\!\begin{pmatrix}
1\\0
\end{pmatrix}
\,,~
v_5\!=\!\begin{pmatrix}
-1\\0
\end{pmatrix}
\,,~
v_6\!=\!\begin{pmatrix}
0\\1
\end{pmatrix}
\,,~
v_7\!=\!\begin{pmatrix}
1\\-1
\end{pmatrix}
\right\}
\, .
\end{align}
The fiber polytope is that of a $\mathbbm{P}^2$ and will give rise to a genus one fibration, and the base is that of a dP$_1$. The corresponding polytopes are given in Figure~\ref{fig:ToricDiagramsP2} and~\ref{fig:ToricDiagramsdP1}, respectively .

\textbf{Fibration 2.} We can also cut the polytope along the hyperplanes with normal directions $(1,0,0,0)$ and $(0,1,0,0)$, i.e.\ the fiber is given by the last two coordinate entries and the base by the first two entries:
\begin{align}
E&\sim\left\{v_1\!=\!\begin{pmatrix}
0\\0
\end{pmatrix}
\,,~
v_2\!=\!\begin{pmatrix}
0\\0
\end{pmatrix}
\,,~
v_3\!=\!\begin{pmatrix}
0\\0
\end{pmatrix}
\,,~
v_4\!=\!\begin{pmatrix}
1\\0
\end{pmatrix}
\,,~
v_5\!=\!\begin{pmatrix}
-1\\0
\end{pmatrix}
\,,~
v_6\!=\!\begin{pmatrix}
0\\1
\end{pmatrix}
\,,~
v_7\!=\!\begin{pmatrix}
1\\-1
\end{pmatrix}
\right\}
\,,\nonumber\\
B&\sim\left\{v_1\!=\!\begin{pmatrix}
1\\0
\end{pmatrix}
\,,~
v_2\!=\!\begin{pmatrix}
0\\1
\end{pmatrix}
\,,~
v_3\!=\!\begin{pmatrix}
-1\\-1
\end{pmatrix}
\,,~
v_4\!=\!\begin{pmatrix}
0\\0
\end{pmatrix}
\,,~
v_5\!=\!\begin{pmatrix}
0\\0
\end{pmatrix}
\,,~
v_6\!=\!\begin{pmatrix}
0\\0
\end{pmatrix}
\,,~
v_7\!=\!\begin{pmatrix}
0\\0
\end{pmatrix}
\right\}
\, .
\end{align}
Now, the fiber polytope is that of dP$_1$ and will give rise to an elliptic fibration with two sections, while the base is a $\mathbbm{P}^2$. The toric diagrams are given in Figure~\ref{fig:ToricDiagramsdP1} and~\ref{fig:ToricDiagramsP2}, respectively. 

Note that the third fibration is not realized torically, but can be found using Kollar's criterion as explained above.

\subsection{K3 fibrations}
\label{sec:K3Fib}
The way in which K3 fibrations are encoded in the enhancement diagrams is very similar to the elliptic case. We observe the following correspondence:
\begin{enumerate}
\item The total number of K3 fibrations is given by the number of Type II$_b$ vertices in the enhancement diagram
\end{enumerate}
According to Oguiso~\cite{Oguiso:1993aaa}, all these fibrations have a $\mathbbm{P}^1$ base. Hence the case of birationally equivalent bases we encountered for genus one fibrations does not occur. This also means that it can never happen that two Type II$_b$ vertices are connected.

\subsubsection{Identifying K3 fibrations}

In order to find the number of K3 fibrations, we use very similar techniques to the ones for elliptic fibrations. The only difference is that we identify two-dimensional rather than one-dimensional fibers.

For the toric CY constructions, we use \texttt{PALP} to find reflexive codimension one subpolytopes that define the K3 fiber, i.e.\ we identify toric K3 fibrations, where the fibration is inherited from the ambient space polytope. For the CICYs, we also concentrate on the K3 fibrations inherited from the ambient space in an obvious way, following~\cite{Anderson:2017aux}. The prescription is the same as in~\eqref{eq:OGF}, except that the CICY defining the fiber should be two-dimensional rather than one-dimensional.

\subsubsection{Example: Enhancement diagrams of K3 fibered CYs}
\begin{figure}[t]
\centering
\includegraphics[height=5cm]{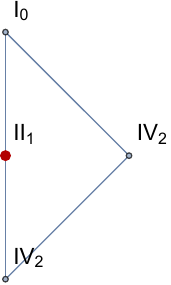}\hspace*{1cm} 
\includegraphics[height=5cm]{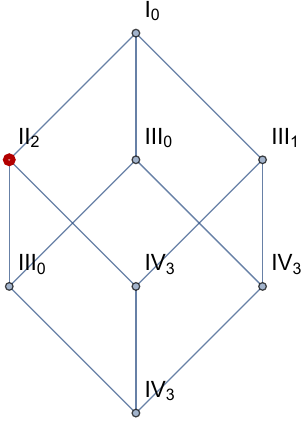}\hspace*{1cm} 
\includegraphics[height=5cm]{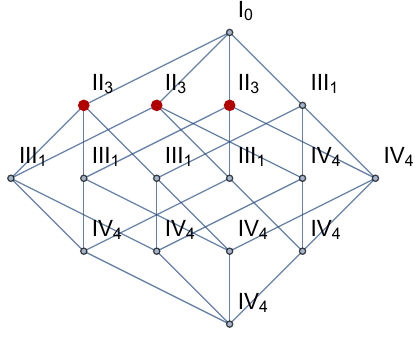}\hspace*{1cm} 
\caption{Three example enhancement diagrams to identify K3 fibrations, which are given by vertices of degeneration Type II.}
\label{fig:graphcountK3}
\end{figure}

We present three examples of K3-fibered CYs in Figure~\ref{fig:graphcountK3}. They are obtained from the following CYs:
\begin{enumerate}
\item The first diagram corresponds to CICY 7806 or KS polytope 34 with triangulation
\begin{align}
\text{SR}=\langle z_1 z_2,\,  z_0 z_3 z_4 z_5 \rangle\;.
\end{align}
\item The second diagram is the same one we used to discuss elliptic fibrations in Section~\ref{sec:EllFib}, i.e.\ CICY 7875 or KS polytopes 103 or 111.
\item The third diagram corresponds to CICY 7859 or KS polytopes 288, 643, or 668 (the polytopes have a unique fine regular star triangulation).
\end{enumerate}

By counting the number of Type II vertices, we find that the three examples have 1, 1, and 3 K3 fibrations, respectively. In order to illustrate the procedure of identifying these, we use the example CY we already discussed in Section~\ref{sec:ExampleEllFib}. We repeat the diagram in Figure~\ref{fig:graphcountK3}, but this time we highlight the Type II vertex.

For the CICY representation of the manifold, we need to find a block that defines a complex two-dimensional fiber. This fiber is given by using row one and three, and both columns from the configuration matrix~\eqref{eq:ellFibExampleCICY},
\begin{align}
\label{eq:K3FibExampleCICY}
\text{K3}\sim\left[
\begin{array}{c|cc}
\mathbbm{P}^2&0&3\\
\mathbbm{P}^2&1&2
\end{array}
\right]\;,\qquad
B\sim \left[
\begin{array}{c}
\mathbbm{P}^1
\end{array}
\right]\;,
\end{align}
and the base is just the $\mathbbm{P}^1$ ambient space factor, as was to be expected for a K3 fibration.

We can also see the K3 fibration torically. In order to identify it, we need to find a single codimension one hypersurface along which to cut the 4D reflexive polytope. In the example at hand, this is given by the hyperplane with normal vector $(0,0,0,1)$. We then find for the fiber a K3 polytope and for the base again a $\mathbbm{P}^1$,
\begin{align}
E&\sim\left\{v_1\!=\!\begin{pmatrix}
1\\0\\0
\end{pmatrix}
\,,~
v_2\!=\!\begin{pmatrix}
0\\1\\0
\end{pmatrix}
\,,~
v_3\!=\!\begin{pmatrix}
-1\\-1\\0
\end{pmatrix}
\,,~
v_4\!=\!\begin{pmatrix}
0\\0\\1
\end{pmatrix}
\,,~
v_5\!=\!\begin{pmatrix}
0\\0\\-1
\end{pmatrix}
\,,~
v_6\!=\!\begin{pmatrix}
0\\0\\0
\end{pmatrix}
\,,~
v_7\!=\!\begin{pmatrix}
0\\0\\1
\end{pmatrix}
\right\}
\, ,\nonumber\\
B&\sim\left\{v_1\!=\!\begin{pmatrix}
0
\end{pmatrix}
\,,~~~~
v_2\!=\!\begin{pmatrix}
0
\end{pmatrix}
\,,~~~~
v_3\!=\!\begin{pmatrix}
0
\end{pmatrix}
\,,~~~~
v_4\!=\!\begin{pmatrix}
0
\end{pmatrix}
\,,~~~~
v_5\!=\!\begin{pmatrix}
0
\end{pmatrix}
\,,~~~~
v_6\!=\!\begin{pmatrix}
1
\end{pmatrix}
\,,~
v_7\!=\!\begin{pmatrix}
-1
\end{pmatrix}
\right\}
\, .
\end{align}
The corresponding toric diagrams are given in Figure~\ref{fig:ToricDiagramsK3}

\subsection{Nested fibrations}
\label{sec:NestedFib}
Often, CYs are K3 fibered, and the K3 itself is elliptically fibered over a $\mathbbm{P}^1$. We observe the following correspondence:
\begin{enumerate}
\item A nested fibration corresponds to an edge that connects a Type II vertex to a Type III vertex. The total number of nested fibrations is then given by the number of such edges.
\end{enumerate}
Note that the same elliptic fibration can appear in different K3 fibrations, i.e.\ a Type III vertex can have edges connecting them with multiple Type II fibers.

\subsubsection{Identifying nested fibrations}
In order to find nested fibrations, we can simply combine the two techniques explained above to find K3 and elliptic fibrations, respectively. Again, we use \texttt{PALP} and the results of~\cite{Anderson:2017aux} for the KS and CICY case, respectively.

\subsubsection{Example: Enhancement diagrams of nested fibrations}
\begin{figure}[t]
\centering
\includegraphics[height=5cm]{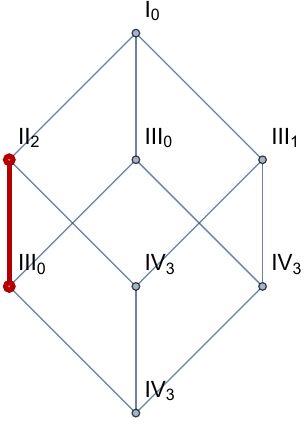}
\caption{An example enhancement diagrams to identify nested fibrations, i.e.\ K3 fibrations which are also elliptically fibered. The nesting is given by edges connecting Type II to Type III vertices.}
\label{fig:graphcountNested}
\end{figure}

We present our working example in Figure~\ref{fig:graphcountNested}, i.e.\ CICY 7875 or KS polytopes 103 or 111. This diagram has a Type II vertex which is connected to a Type III vertex. The corresponding nested fibration is given by the K3 identified in Section~\ref{sec:K3Fib}, whose elliptic fiber corresponds to fibration 1 in Section~\ref{sec:EllFib}.

In more detail, we can see that for the CICY, the elliptic fibration~\eqref{eq:ellFibExampleCICY} also sits inside the K3 fibration given by the CICY in~\eqref{eq:K3FibExampleCICY}, i.e.\ the elliptic curve is given by a cubic in $\mathbbm{P}^2$, and the base $\mathbbm{P}^1$ of the fibration is given by a linear in $\mathbbm{P}^2$, which defines a $\mathbbm{P}^1$.

Similarly for the toric case, we can find the elliptic fiber in the first three vertices of the K3 fiber, by cutting in addition along the hypersurface with normal direction $(0,0,1)$ inside the three-dimensional polytope whose anticanonical hypersurface defines the K3. Coordinates four and five can be seen to define a $\mathbbm{P}^1$, and the last two correspond to the origin, i.e.\ the unique interior point of all nested subpolytopes. In the toric diagram of Figure~\ref{fig:ToricDiagramsK3}, we can also see the nested structure by noticing that the 2D polytope corresponding to the fiber (the polytope of the ambient $\mathbbm{P}^2$ in Figure~\ref{fig:ToricDiagramsP2}) is at height 0, while the 2D polytope that defines the base $\mathbbm{P}^1$ (cf.~Figure~\ref{fig:ToricDiagramsP1}) is given by the origin plus the two vertices at height~$\pm1$.

%%%%%%%%%%%%%%%%%%%%%%%%%%%%%%%%%%%%%%%%%%%%%%%%%%%%%%%%%%%%%%%%%%%%%%%%%%%%
\section{Classification of CY threefolds and statistics}
\label{sec:Classification}
\begin{table}[t]
\centering
\begin{tabular}{| c || c | c | c | c | c |}
\hline 
 & $h^{1,1}=1$ & $h^{1,1}=2$ & $h^{1,1}=3$ & $h^{1,1}=4$ & $h^{1,1}=5$ \\
\hline \hline
Geometries & 5 & 48 & 393 & 2536 & 17411\\
Distinct diagrams & 1 & 6 & 32 & 209 & 950 \\ \hline
\end{tabular}
\caption{Statistics for the Kreuzer-Skarke examples, including only simplicial K\"ahler cones.}
\label{table:KS}
\end{table}

\begin{table}[t]
\centering
\begin{tabular}{| c || c | c |}
\hline 
& $h^{1,1}=3$ & $h^{1,1}=4$ \\
\hline \hline
Geometries & 58 & 1399 \\
Distinct diagrams & 6 & 165 \\ \hline
\end{tabular}
\caption{Statistics for Kreuzer-Skarke examples, including only non-simplicial K\"ahler cones.}
\label{table:KSnon-simpl}
\end{table}

\begin{table}[t]
\centering
%footnotesize
\footnotesize
\begin{tabular}{| c || c | c | c | c | c | c | c | c | c | c |}
\hline 
 & \!\!$h^{1,1}=1$ \!\!\!&\!\! $h^{1,1}=2$ \!\!\!&\!\! $h^{1,1}=3$  \!\!\!&\!\! $h^{1,1}=4$  \!\!\!&\!\! $h^{1,1}=5$  \!\!\!&\!\! $h^{1,1}=6$  \!\!\!&\!\! $h^{1,1}=7$  \!\!\!&\!\! $h^{1,1}=8$  \!\!\!&\!\! $h^{1,1}=9$  \!\!\!&\!\! $h^{1,1}=10$ \!\!\!\\
\hline \hline
Geometries & 5 & 36 & 155 & 425 & 856 & 1257 & 1462 & 1325 & 1032 & 643\\
Diagrams & 1 & 4 & 11 & 25 & 53 & 117 & 235 & 220 & 271 & 267 \\ \hline
\end{tabular}
\caption{Statistics for the CICY examples.}
\label{table:CICY}
\end{table}

\begin{table}[t]
\centering
\begin{tabular}{| c || c | c | c | c | c |}
\hline 
 & $h^{1,1}=1$ & $h^{1,1}=2$ & $h^{1,1}=3$ & $h^{1,1}=4$ & $h^{1,1}=5$ \\
\hline \hline
Distinct diagrams & 1 & 6 & 34 & 219 & 950\\ \hline
\end{tabular}
\caption{Statistics for combined Kreuzer-Skarke and CICY scan, including only simplicial K\"ahler cones.}
\label{table:KS_CICY}
\end{table}
Here we discuss how we can tell Calabi-Yau threefolds apart from each other via their enhancement diagrams, and how these graphs thus serve as a means to classify CYs. To exhibit the use of enhancement diagrams for such a classification, we discuss the results obtained in our scans of the Kreuzer-Skarke and CICY data sets.  The numbers of distinct enhancement diagrams found in each of these scans have been listed in Tables \ref{table:KS}, \ref{table:KSnon-simpl}, \ref{table:CICY} and \ref{table:KS_CICY}, and the diagrams obtained up to $h^{1,1}=3$ are provided in Appendix \ref{app:diagrams}. 

By Wall's theorem \cite{Wall:1966aaa}, we find that two CYs are homotopically inequivalent when their enhancement diagrams are inequivalent, since it implies that their triple intersection numbers cannot be related to each other via a basis transformation for $H^2(Y_3)$. In this sense the enhancement diagrams serve as an invariant that can be computed for each Calabi-Yau threefold, and can be used to determine whether two Calabi-Yau threefolds are the same. Note however that these diagrams are not yet fine enough to distinguish all CYs,  since we obtain for instance the same diagram for all CYs with $h^{1,1}=1$, because they only have a Type IV$_1$ degeneration. Wall's theorem suggests that we would need to incorporate other topological data such as the second Chern class $c_2(Y_3)$ of the Calabi-Yau threefold into the diagram. Note that while the integrated second Chern classes $c_I$ appear explicitly in the definition of the nilpotent orbit in $\mathbf{a}_0$~\eqref{nil-gen} and consequently also in the definition of the limiting mixed Hodge structure, the enhancement diagrams do not explicitly depend on them. It would be interesting to refine the classification to capture this information as well.

From a practical perspective, extending this analysis to larger values of $h^{1,1}$ becomes rather tedious, since the number of vertices grows as $h^{1,1}!$.  Therefore one can resort to more pragmatic checks first when comparing enhancement diagrams, as was already touched upon briefly in Section \ref{sec:symmetries}. For instance, one could simply count how often each degeneration type occurs in the diagram, and these numbers then serve as invariants extracted from the diagram.  Note that such extracted invariants lead to a less refined classification then the diagrams themselves, as demonstrated by the two distinct diagrams in Figure \ref{fig:hassediagramsymmetries} which consist of the same numbers of degeneration types. Nevertheless these numbers would capture properties of the Calabi-Yau manifold, since --- in the spirit of Section \ref{sec:DiagramsFibrations} --- they would capture to the number of elliptic fibrations (number of III vertices) and K3 fibrations (number of II vertices). For low values of $h^{1,1}$ comparing graphs is not an issue, cf. Tables \ref{table:KS}, \ref{table:KSnon-simpl}, \ref{table:CICY} and Appendix \ref{app:diagrams}.%Following Section \ref{sec:symmetries}, we can refine the invariants by instead counting sets of indistinguishable vertices at each row of the enhancement diagram, including sets that consist of a single vertex. This actually leads to the same classification for our scan of the Kreuzer-Skarke database as the graphs. It results in a set of numbers $n^{r,v}_{\text{A}}$, where A indicates the degeneration type, $r$ the row of the diagram, and $v$ the number of vertices in this set of indistinguishable vertices.\Dv{Probably do not want to introduce these numbers $n^{r,v}_{\text{A}}$, better to just convey the idea of extracting suitable invariants, especially with the next paragraph in mind} 

To demonstrate the use of enhancement diagrams to classify Calabi-Yau threefolds, we next discuss the results that were obtained by scanning the Kreuzer-Skarke and CICY data sets. First of all, note that there is a larger number of distinct diagrams found from the KS data set than from the CICY data set, as can be seen from Tables \ref{table:KS} and \ref{table:CICY}, although CICY does provide some new diagrams as follows from Table \ref{table:KS_CICY}. Furthermore, there is some overlap in the set of diagrams obtained from the two scans, which follows partly from certain Calabi-Yau threefolds being present in both sets, such as e.g.~the quintic. On the one hand, this smaller number of distinct diagrams in the CICY scan follows simply from the fact that the Kreuzer-Skarke data set has a larger number of CYs for every value of $h^{1,1}$. Another explanation for this lack of variation in the CICY scan follows from the subindex of the IV vertices that occur in the diagrams. Namely, for CICY threefolds we find that every Type IV limit has subindex $h^{1,1}$, i.e.~all IV vertices in their enhancement diagrams are IV$_{h^{1,1}}$ vertices, %\footnote{We believe this to be related to their ambient spaces, which are products of projective spaces, and the fact that their K\"ahler cone descends from these ambient spaces.} 
whereas we did not observe such restrictions for the Kreuzer-Skarke database.

%%%%%%%%%%%%%%%%%%%%%%%%%%%%%%%%%%%%%%%%%%%%%%%%%%%%%%%%%%%%%%%%%%%%%%%%%%%%
\section{Conclusion and outlook}
\label{sec:Conclusion}

In this paper we introduced a new way to classify Calabi-Yau threefolds by using asymptotic Hodge theory. 
More concretely, we suggest that to any decompactification limit performed in the K\"ahler moduli space 
of $Y_3$ we can associate a limiting mixed Hodge structure. 
These structures are defined by the large volume monodromy transformations $N_i$, which are given in terms of 
the triple intersection numbers, and a limiting vector $\mathbf{a}_0$, defined in terms of the integrated Chern classes of $Y_3$. 
Of key importance in this work has been the fact that the data $(N_i,\mathbf{a}_0)$ does not only define a single limiting 
mixed Hodge structure, but rather a collection of such structures, one for each ordered limit in K\"ahler moduli 
space obtained by sending step-wise volume moduli to infinity. We have combined this property with 
the recent classification of all possible types of limiting mixed Hodge structures arising for Calabi-Yau threefolds
in K\"ahler and complex structure moduli space \cite{Kerr2017,Grimm:2018cpv,Corvilain:2018lgw}. For an $m$-dimensional moduli space these structures are 
categorized into $4m$ degeneration types denoted by I$_a$, II$_b$, III$_c$, IV$_d$. We collected the types arising at all 
possible decompactification limits in an enhancement diagram. This provided us with a Hasse diagram that starts 
at the non-degenerate case I$_0$ and ends on the maximal degeneration IV$_{h^{1,1}(Y_3)}$. We then 
argued that these enhancement diagrams provide a classification of all Calabi-Yau threefolds capturing many 
of their intrinsic properties.

It is important to stress the enhancement diagrams are not one-to-one with homotopically distinct  
Calabi-Yau manifolds. We have seen this in various different ways. Firstly, we have argued that the degeneration 
types in the K\"ahler moduli space only depended on the intersection numbers, and hence not all 
data distinguishing the homotopy types of Calabi-Yau threefolds according to Wall's theorem \cite{Wall:1966aaa}. 
Secondly, we have computed the enhancement diagrams for large sets of examples taken from the 
Kreuzer-Skarke list \cite{Kreuzer:2000xy} and CICY list \cite{Candelas:1987kf,Anderson:2017aux} of Calabi-Yau threefolds. We then compare the 
number of diagrams to the number of Calabi-Yau threefolds indicating how many fall into the equivalence 
classes represented by a distinct enhancement diagram. This was generally possible and included cases 
with both simplicial and non-simplicial K\"ahler cones. The enhancement diagrams turned out to have a rich structure 
and the number of possibilities grows when increasing $h^{1,1}(Y_3)$. Therefore, they provide 
a powerful tool to decide when two threefolds are actually homotopically different and to analyze some of its 
important topological features. 

Note that all limits in the K\"ahler cone lead to a decompactification of the Calabi-Yau manifold.
The classification into types II$_a$, III$_b$, IV$_c$ captures the information about the type of submanifolds that grow in a considered 
possibly multi-variable limit. This should be contrasted with the equi-dimensional limits studied in \cite{Lee:2019wij}, where 
the emergence of weakly coupled strings was proposed. These 
limits require to approach the walls of the K\"ahler cone at which the volume of the Calabi-Yau stays finite. In Type II
compactifications such limits receive quantum corrections which are under control using mirror symmetry. 
One then can use the strategy presented in this work and determine an associated enhancement pattern. 
It is crucial, however, to not use the large volume monodromies \eqref{lcsNa} determined in terms of the intersection numbers, but 
rather the monodromy associated to this specific limit using the same classification method \cite{Kerr2017,Grimm:2018cpv,Corvilain:2018lgw}. 

Based on studying Calabi-Yau threefolds constructed as hypersurfaces in a 4D toric ambient space as well as from complete intersections in an ambient space that is given by a product of projective spaces, we have identified how fibration structures are encoded in the enhancement diagrams. Since fibrations lead to special intersection patterns, the presence of fibrations is encoded in the triple intersection numbers: . For example, as we explained in Section~\ref{ssec:EllipticFibrationsKollar}, a divisor that is located purely in the base has vanishing triple intersection numbers, while it's double intersection leads to a non-trivial curve in the base. As classified by Oguiso, a Calabi-Yau threefold can have a fibration where the fiber is either a $T^2$, a K3, or a $T^4$. Moreover, there can be nested fibrations, i.e.\ the K3 can itself be elliptically fibered over $\mathbbm{P}^1$. All this information can be read off directly from the enhancement diagrams without the need to identify the base, the fiber, a section, the Kollar divisor, etc. The rules are very simple:
\begin{enumerate}
\item Genus one fibrations are in one-to-one correspondence with type III vertex in the enhancement diagram. Birationally equivalent fibrations are connected by edges.
\item K3 fibrations are in one-to-one correspondence with type II vertices  in the enhancement diagram. The base is always a $\mathbbm{P}^1$.
\item Calabi-Yau manifolds that are K3 fibered, where the K3 itself is elliptically fibered have a type II and a type III vertex. The nested fibration is encoded in an edge that connects the two vertices.
\end{enumerate}

One of the most interesting future directions is to use our proposed classification of Calabi-Yau manifolds 
as a guiding principle to classify supergravity theories. In particular, compactifying M-theory on 
a Calabi-Yau threefold yields a five-dimensional minimally supersymmetric supergravity theory. 
In this compactification the triple intersection numbers determine the vector multiplet metric and Chern-Simons couplings 
while the integrated second Chern class fixed certain higher-curvature terms. As discussed in this paper the same 
data is crucial in fixing a collection of  mixed Hodge structures arising at all possible limits in K\"ahler moduli space. 
It is then suggestive to bypass the step of having a geometric compactification and 
directly specify the data of the vector sector of every five-dimensional supergravity theory by a collection 
of a certain type of mixed Hodge structures. The rich set of consistency conditions arising from the underlying deep mathematical 
structure might yield unexpected constraints that allow identifying models that are in the swampland of five-dimensional 
supergravity theories. In particular, we believe that the arising constraints are stronger then known 
consistency conditions, such as the requirement of having positive kinetic terms. 

Another very interesting question is to use this classification to study transitions in Calabi-Yau geometries. Equipped with an understanding of what type of transitions can occur purely on the level of the diagrams could shed light on the moduli space of Calabi-Yau threefolds. As we have seen in Section \ref{sec:recursion}, the first three rows of the enhancement diagrams fix all possible indices for the degeneration types of subsequent rows, thus putting a lower bound on the number of inequivalent Calabi-Yau threefolds. In addition, we have exemplified in Section~\ref{sec:QuinticExample} that, starting from the quintic, one can obtain more complicated diagrams by including blowups that lead to conifold or del Pezzo transitions. Moreover, both blowups can be performed simultaneously, which results in a combination of the diagrams of the individual cases. This begs the question whether the connectedness of simply-connected Calabi-Yau threefolds, sometimes referred to as Reid's fantasy~\cite{Reid:1987aaa}, can be studied in a diagrammatic way, purely using the types of allowed transitions. One could also turn the question around and ask which types of enhancement or transition rules would allow for a finite number of Calabi-Yau threefolds. In fact, a simpler question might be the following: given the fact that the number of elliptically fibered Calabi-Yau threefolds is finite~\cite{Grassi:1991aaa,Gross:1994aaa}, what are the corresponding constraints on diagrams (or equivalently mixed Hodge structures) with type III degenerations that lead to a finite number of possible diagrams? We leave these interesting open questions for future work.

Lastly, identifying the correspondence between edges and vertices of a certain type on the one hand and (nested or birationally equivalent) fibrations on the other was done by studying a large class of examples in this paper. However, the diagrammatic way of representing properties of Calabi-Yau threefolds lends itself to studies via data science and machine learning, a growing subfield in the analysis of string theory~\cite{He:2017aed,Krefl:2017yox,Ruehle:2017mzq,Carifio:2017bov}. For example, supervised machine learning has been used previously to classify whether CICYs are elliptically fibered, based on their intersection numbers~\cite{He:2019vsj}. Other Calabi-Yau or physics data, such as the gauge group or the spectrum that arises e.g.\ from compactification of F-Theory on these manifolds, is likely encoded in the enhancement diagrams as well. These connections could be identified with white-box supervised machine learning techniques such as decision trees, or via unsupervised techniques such as feature extraction. Moreover, it is interesting to study whether this graph structure helps with searches in the Kreuzer-Skarke database for models with specific properties (such as finding threefolds with specific fibrations or gauge groups), using supervised learning with deep or graph neural networks.

\subsection*{Acknowledgments}
It is a pleasure to thank Pierre Corvilain, Markus Dierigl, Amihay Hanany, Yang-Hui He, Seung-Joo Lee, Chongchuo Li, Eran Palti, and Irene Valenzuela for helpful discussions. F.R.~thanks Utrecht University for hospitality during the final stage of this project.

\appendix

%%%%%%%%%%%%%%%%%%%%%%%%%%%%%%%%%%%%%%%%%%%%%%%%%%%%%%%%%%%%%%%%%%%%%%%%%%%%
\section{Limiting mixed Hodge structures} \label{limitingMHS}
In this appendix we briefly introduce the mathematical notion of a limiting mixed Hodge structure. 
It should be stressed, however, that our exposition is short and incomplete. We refer the 
reader to the original papers \cite{Schmid,CKS} and the review \cite{CKAsterisque}. Also \cite{Grimm:2018cpv} contains 
a concise summary of some of the relevant aspects. 

Let us first define a \textit{pure} Hodge structure and its associated Hodge filtration. Let $V$ be a rational vector space.
A pure Hodge structure of weight $w$ describes a decomposition of the complexification $V_{\mathbb{C}} = V \otimes \mathbb{C}$ 
as
\beq \label{Hodge-decomp_app}
   V_\mathbb{C} = \cH^{w,0}  \oplus  \cH^{w-1,1}  \oplus \ldots  \oplus  \cH^{1,w-1}  \oplus \cH^{0,w}\; ,
\eeq
with the subspaces satisfying $\cH^{p,q} = \overline{\cH^{q,p}}$ with $w=p+q$. The complex conjugation on $V_\mathbb{C}$ 
is defined with respect to the rational vector space $V$. The $\cH^{p,q}$ can also be used  
to define a Hodge filtration by setting $F^p = \oplus_{i\geq p} \cH^{i,w-i}$. These spaces are filtered and satisfy  
\beq \label{F-filtration}
   V_\mathbb{C} = F^0 \ \supset \ F^1 \ \supset\  \ldots\ \supset\   F^{w-1}\ \supset\   F^w = \cH^{w,0}\; ,
\eeq
such that $\cH^{p,q} = F^p \cap \bar F^q$. A crucial additional property arises from demanding that the 
$\cH^{p,q}$ define a
 \textit{polarized} pure Hodge structure. This necessitates the 
 existence of a bilinear form $S(\cdot, \cdot)$ on $V_\mathbb{C}$, such that the conditions  
\bea
     S(\cH^{p,q}, \cH^{r,s}) &=& 0\; , \qquad   p \neq s,\ q \neq r\; , \\
     i^{p-q} S(v,\bar v) &>& 0\; ,  \qquad \text{for all}\ \ 0\neq v \in \cH^{p,q}\; ,
\eea
are satisfied. Note that these definitions define a fixed $(p,q)$-splitting. 
One can then ask the question how such a structure can vary consistently over a complex 
base space $\cM$ and define 
families of polarized pure Hodge structures. This is captured by the theory of 
variation Hodge structures. In particular, one demands that with respect to the flat connection $\nabla$
 on the family of Hodge structures varying over $\cM$, 
 the $F^i$ are holomorphic sections and satisfy $\nabla F^i \subset F^{i-1} \otimes \Omega_\cM$. 
Let us note that these definitions are essentially algebraic. The application to 
geometric settings arises, for example, when using them to describe the $(p,q)$-cohomology $H^{p,q}(Y_3,\mathbf{C})$
of a Calabi-Yau threefold $Y_3$ and how it varies over the complex structure moduli space. 

Let us next turn to the definition of a mixed Hodge structure. The crucial new ingredient is the 
so-called \textit{monodromy weight filtration} $W_i$. This filtration is induced by the action 
of a nilpotent matrix $N$ on the vector space $V$. Concretely, one defines the 
rational vector subspaces $W_{j} (N) \subset 
V$ by requiring that they form a filtration
\beq 
W_{-1}\equiv 0\ \subset\  W_0\ \subset\ W_1\ \subset\ ...\  \subset\ W_{2w-1}\  \subset \ W_{2w} = V\; ,
\label{filtration}
\eeq
with the properties
\begin{align}
 &  1.) \quad N W_i \subset W_{i-2} \; ,&\\
   & 2.) \quad  N^j : Gr_{w+j} \rightarrow Gr_{w-j}\ \ \text{is an isomorphism,}\quad Gr_{j} \equiv W_{j}/W_{j-1} \; .& \label{iso-prop}
\end{align} 
Here the quotients $Gr_i$ are equivalence classes of elements of $W_i$ that differ by elements of $W_{i-1}$. One 
can show that the filtration $W_i$ with the above properties is unique for a given $N$.

Finally, let us define a \textit{mixed Hodge structure} $(V,W,F)$. Let $W_i$ be a monodromy weight filtration defined 
by an $N$ as above and $F^q$ a filtration satisfying \eqref{F-filtration} on the vector space $V$.\footnote{Note that 
the $F^p$ do not need to define a pure Hodge structure.} We require that $N$ is compatible 
with the $F^q$-filtration and acts on it horizontally, i.e.~$N F^p \subset F^{p-1}$.
 The defining feature of a mixed Hodge structure is that each $Gr_{j}$ defined in \eqref{iso-prop} 
admits an induced Hodge filtration 
\beq \label{mixedHodge-filtration}
    F^p Gr_j^\mathbb{C} \equiv ( F^p  \cap W_j^\mathbb{C} )/ ( F^p  \cap W_{j-1}^\mathbb{C})\; ,
\eeq
where $Gr_j^\mathbb{C} = Gr_j \otimes \mathbb{C}$ and $W_i^\mathbb{C} = W_i \otimes \mathbb{C}$ are the complexification.
Referring back to \eqref{Hodge-decomp_app}, this implies that we can split each $Gr_j $ into a pure Hodge structure $\cH^{p,q}$
as  
\beq \label{Grj-split}
   Gr_j  = \bigoplus_{p+q=j} \cH^{p,q} \; ,\qquad   \cH^{p,q} =  F^p Gr_j \cap \overline{F^q Gr_j}\; ,
\eeq
where we recall that $w=p+q$ is the weight of the corresponding pure Hodge structure. 
The operator $N$ is a morphism among these pure Hodge structures. 
Using the action of $N$ on $W_i$ and $F^p$, we find $N Gr_j \subset Gr_{j-2}$ and 
$N \cH^{p, q} \subset \cH^{p-1,q-1}$. Note that this induces a jump in the weight of the pure Hodge structure by $-2$, 
while the mixed Hodge structure is preserved by $N$. The natural next step is to introduce a \textit{polarized mixed Hodge structure}. 
This again uses the bilinear form $S(\cdot,\cdot)$. We first define the primitive subspaces $\cP_{l} \subset Gr_{l+w}$, by setting $\cP_{l} = ker(N^{l+1} : Gr_{w+l} \rightarrow Gr_{w-l-2})$. The mixed Hodge structure is polarized if for all $l$ the restriction of the pure Hodge structure \eqref{Grj-split} to the 
primitive subspaces $\cP_l$ is polarized with respect to 
$S_l(\cdot,\cdot) = S(\cdot ,N^l\cdot)$.

With this definition at hand, we can now introduce a limiting mixed Hodge structure. The introduction of this structure is needed due 
to the fact that a pure Hodge structure at certain limits of $\cM$ can degenerate and no longer describe the splitting of $V_{\mathbb{C}}$.
Let us describe a one-parameter degeneration limit $t\rightarrow i \infty$. 
At such a limit one can introduce a nilpotent matrix $N$ from the monodromy transformation as discussed in the main part of the paper. 
One can then split off the singular part of the pure Hodge filtration defining 
\beq
      F_{\infty}^p = \lim_{t\to i \infty} e^{- t N}  F^p \; . 
\eeq 
While the $F_{\infty}^p $ in general do not describe a pure Hodge structure, they can be used to define 
a mixed Hodge structure. This mixed Hodge structure is defined with respect to the limit $t \rightarrow i \infty$ and 
hence known as a limiting mixed Hodge structure. 

%%%%%%%%%%%%%%%%%%%%%%%%%%%%%%%%%%%%%%%%%%%%%%%%%%%%%%%%%%%%%%%%%%%%%%%%%%%%
\section{Enhancement diagrams obtained in KS and CICY scan}
\label{app:diagrams}
\begin{figure}[t]
\vspace*{-2cm}
\centering
\includegraphics[height=1.8cm]{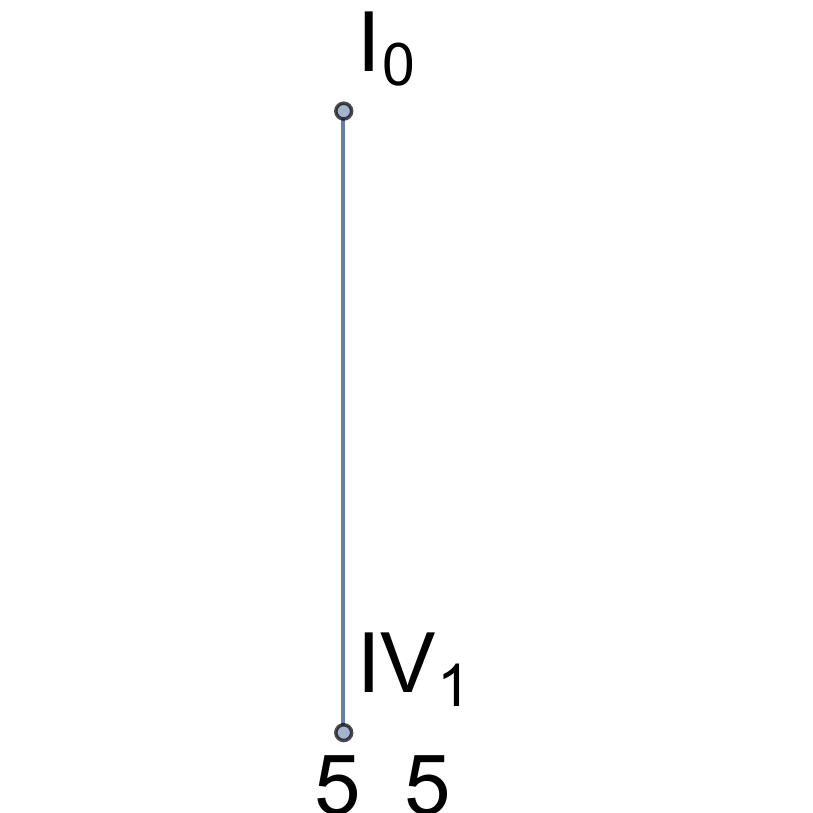}

\includegraphics[width=0.9\textwidth]{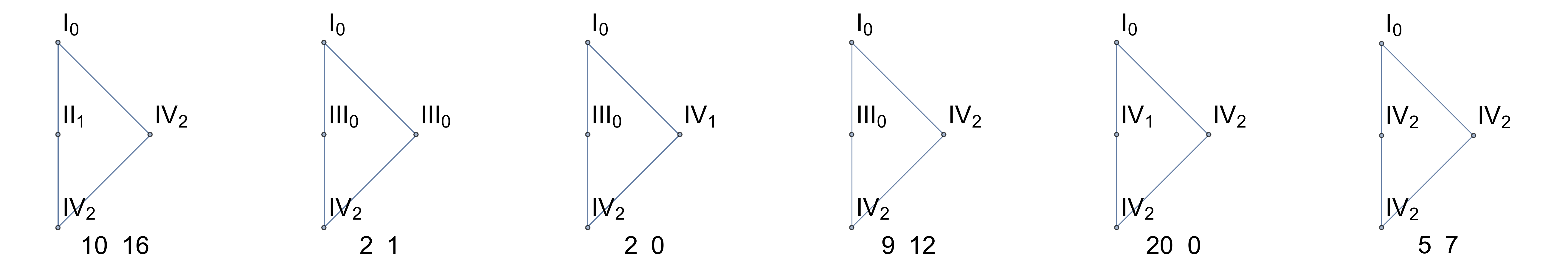}

\includegraphics[width=0.9\textwidth]{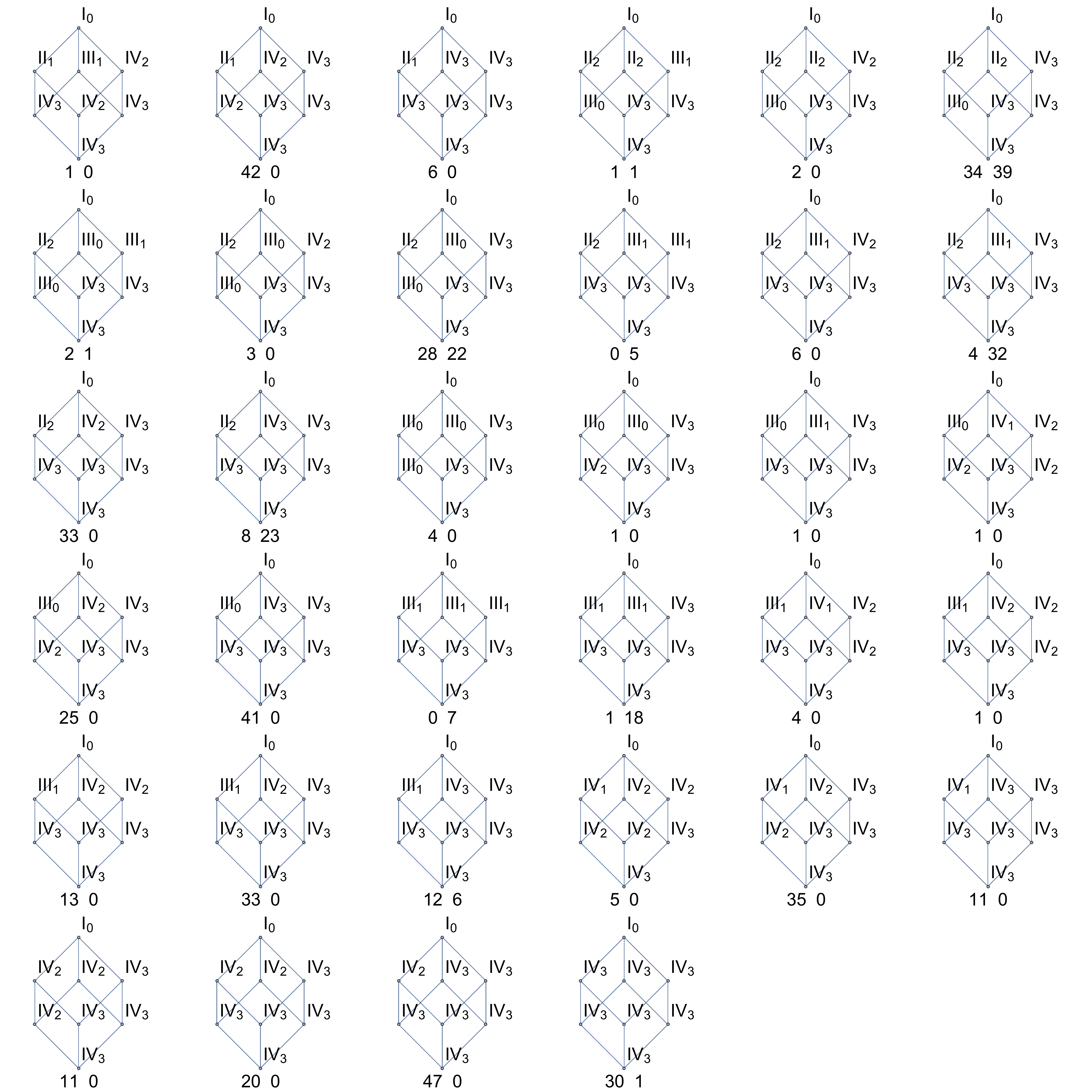}
\vspace{-0.2cm}
\caption{Enhancement diagrams obtained via scans of the Kreuzer-Skarke and CICY databases up to $h^{1,1}=3$, including only simplicial K\"ahler cones. The numbers below each diagram indicate how often it was encountered in the Kreuzer-Skarke and CICY scans respectively.}\label{fig:classification}
\end{figure}

\begin{figure}[t]
\vspace*{-.2cm}
\centering
\includegraphics[width=\textwidth]{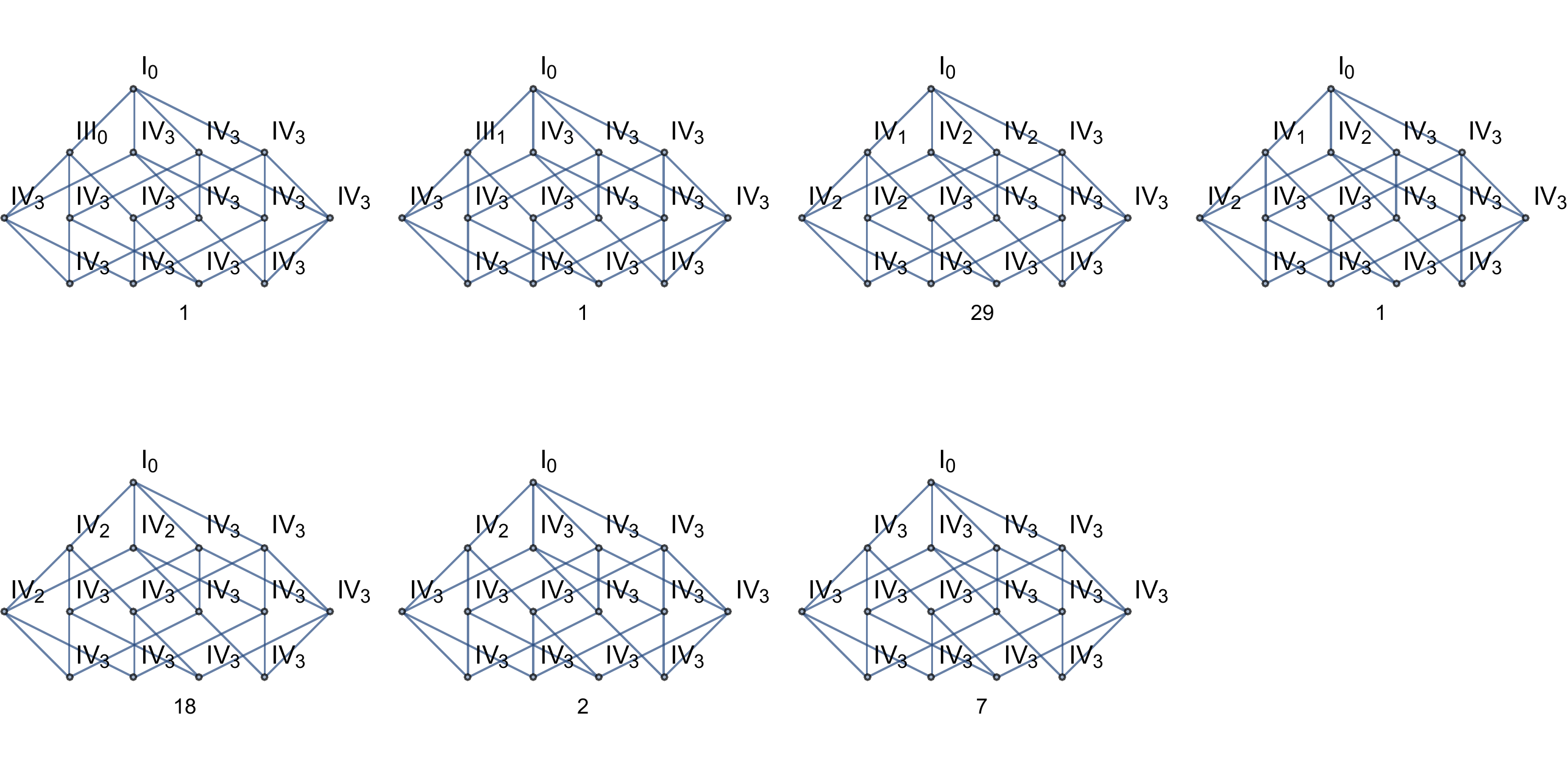}
\vspace{-0.6cm}
\caption{Enhancement diagrams obtained via our scan of the Kreuzer-Skarke database for $h^{1,1}=3$, including only non-simplicial K\"ahler cones, where the number below each diagram indicates its multiplicity in this scan.}\label{fig:classificationNonSimpl}
\end{figure}

We give all distinct diagrams obtained in our scans of the Kreuzer-Skarke and CICY data sets up to $h^{1,1}=3$ for simplicial and non-simplicial K\"ahler cones in Figure~\ref{fig:classification} and \ref{fig:classificationNonSimpl}, respectively. Each diagram is accompanied by the number of times it occurred in both these data sets. 

\clearpage
\bibliographystyle{bibstyle}
\bibliography{refs}
\end{document}